\documentclass[manuscript]{acmart}

\setcopyright{none}
\usepackage{booktabs} 
\usepackage[font=small]{caption}
\usepackage[labelformat=simple]{subcaption}
\usepackage{graphicx}
\usepackage[ruled,vlined,linesnumbered]{algorithm2e}
\usepackage{algorithmic}
\SetKwComment{Comment}{$\triangleright$\ }{}
\usepackage[mathscr]{eucal}
\usepackage{color}
\usepackage{url}
\usepackage{mathtools}
\usepackage{xcolor}
\usepackage{multirow}
\usepackage{soul}
\usepackage{colortbl}

\usepackage{enumitem}
\usepackage[capitalise]{cleveref}

\allowdisplaybreaks

\newcommand{\method}{\textsc{FINEST}\xspace}
\newcommand{\casper}{\textsc{CASPER}\xspace}

\newcommand{\tisasrec}{\textsc{TiSASRec}\xspace}
\newcommand{\lstm}{\textsc{LSTM}\xspace}
\newcommand{\bert}{\textsc{BERT4Rec}\xspace}

\copyrightyear{2024}
\acmYear{2024}
\acmDOI{10.1145/1122445.1122456}
\acmPrice{15.00}
\acmISBN{978-1-4503-XXXX-X/18/06}

\begin{document}

\title{\method: Stabilizing Recommendations by Rank-Preserving Fine-Tuning}

\author{Sejoon Oh}
\email{soh337@gatech.edu} 
\affiliation{%
  \institution{Georgia Institute of Technology}
    \country{United States}
}

\author{Berk Ustun}
\email{berk@ucsd.edu}
\affiliation{%
  \institution{University of California, San Diego}
    \country{United States}
}

\author{Julian McAuley}
\email{jmcauley@eng.ucsd.edu} 
\affiliation{%
  \institution{University of California, San Diego}
    \country{United States}
}

\author{Srijan Kumar}
\email{srijan@gatech.edu} 
\affiliation{%
  \institution{Georgia Institute of Technology}
    \country{United States}
}

\renewcommand{\shortauthors}{Sejoon Oh, Berk Ustun, Julian McAuley, \& Srijan Kumar}

	\begin{abstract}
		\label{sec:abstract}
		\noindent
Modern recommender systems may output considerably different recommendations due to small perturbations in the training data. Changes in the data from a single user will alter the recommendations as well as the recommendations of other users. 
In applications like healthcare, housing, and finance, this sensitivity can have adverse effects on user experience.
We propose a method to stabilize a given recommender system against such perturbations. This is a challenging task due to (1) the lack of  a ``reference'' rank list that can be used to anchor the outputs; and (2) the computational challenges in ensuring the stability of rank lists with respect to all possible perturbations of training data. Our method, \method, overcomes these challenges by obtaining reference rank lists from a given recommendation model and then \emph{fine-tuning} the model under simulated perturbation scenarios with rank-preserving regularization on sampled items. Our experiments on real-world datasets demonstrate that \method can ensure that recommender models output stable recommendations under a wide range of different perturbations without compromising next-item prediction accuracy.

	\end{abstract}

\maketitle
	\section{Introduction}
	\label{sec:intro}
	Modern sequential recommender systems output ranked recommendation lists for users using a model trained from historical user-item interactions~\cite{li2020time, sun2019bert4rec, de2021transformers4rec, kang2018self, hansen2020contextual}. Such recommenders have been widely employed in various applications including E-commerce~\cite{wang2020time, tanjim2020attentive} and streaming services~\cite{hansen2020contextual, beutel2018latent}.

Recent work has shown that  recommendation results generated by
sequential recommenders change considerably as a result of \emph{perturbations} in the training data~\cite{oh2022robustness, yue2021black, betello2023investigating}  -- i.e., changes that would insert, delete, or modify one or more user-item interactions in the training data. In practice, these perturbations can arise due to noise in user-item interactions that are noisy (e.g., a user mistakenly clicking on an item on an online retail website), or as a result of adversarial manipulation~\cite{wu2021triple, zhang2021data} (e.g., a bot creating multiple interactions in a short time).
In the context of recommender systems, this sensitivity can be detrimental to user experience because the data from a single user is used to output recommendations for other users. As a result of this coupling, minor \emph{ interaction-level} perturbations from a single user can lead to drastic changes in the recommendations for \emph{all} users~\citep{oh2022robustness}. In an online platform, this sensitivity could mean that the recommendations for large user segments may change arbitrarily after model retraining. These unexpected changes can reduce user engagement and satisfaction~\cite{jannach2019measuring, pei2019personalized} as users may receive irrelevant items. In extreme cases, an adversary can even intentionally lower the model stability by manipulating the training data, which can amplify user dissatisfaction. In practice, such effects may affect users from certain demographic groups more than others~\cite{oh2022robustness}.

\begin{figure}[t!]
    \includegraphics[width=1.0\linewidth]{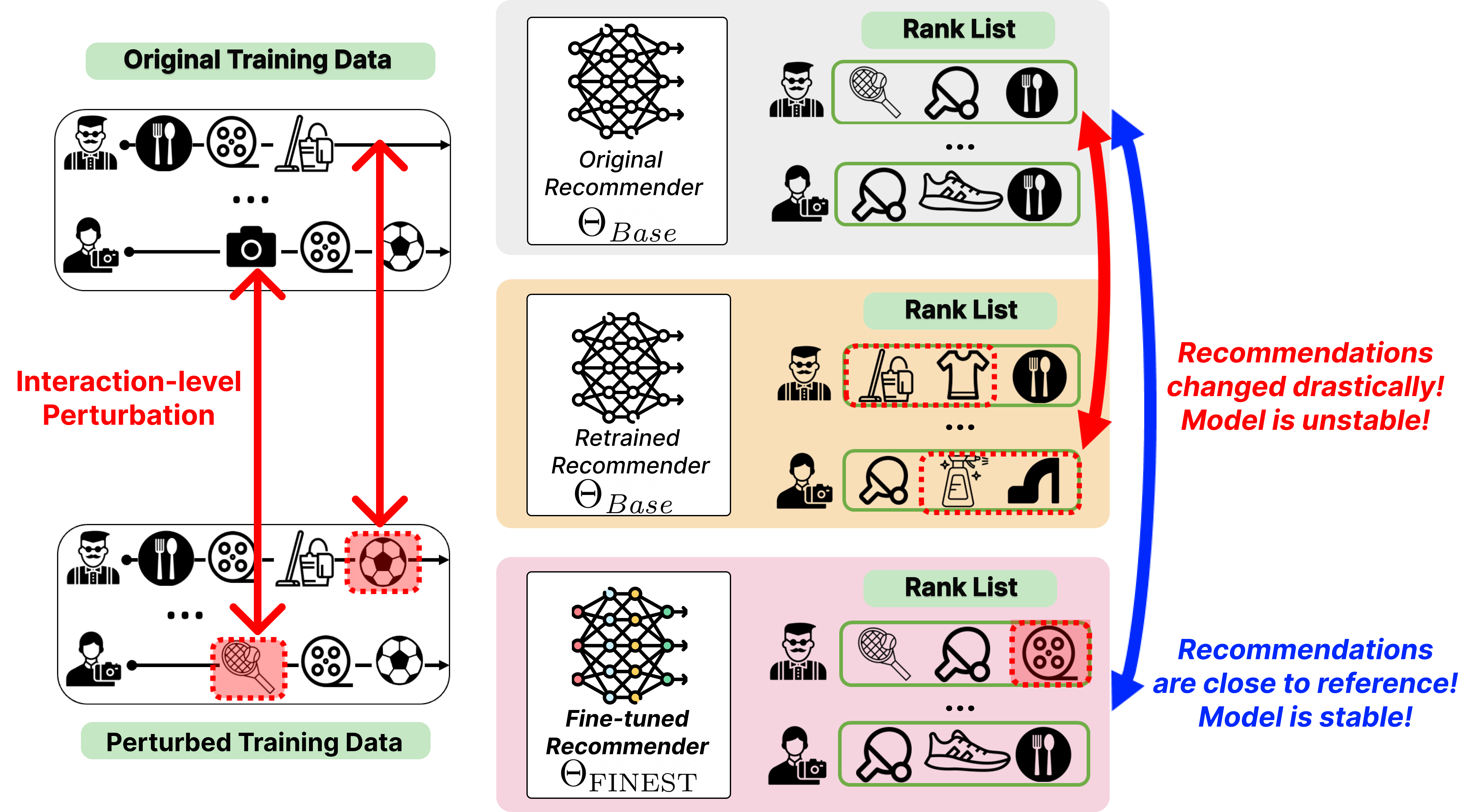}
    \caption{
    Sequential recommendation models can output drastically different rank lists due to small perturbations in user interaction in the training data. Here, we show a training dataset of user interactions (``original") and a copy that contains minor perturbations (``perturbed", with perturbations highlighted in red). Recommendation systems trained using each dataset will output different rank lists for end-users (right-top and  right-middle). Our proposed approach \method (right-bottom) will stabilize outputs to ensure that training with  a ``perturbed dataset" will return rank lists that are as close as possible to the rank list from the ``original dataset."
    }
	\label{fig:intro_plot}
\end{figure}

Despite the importance of recommendation stability, there is limited research on how to induce or enhance stability in recommender systems. Existing methods~\cite{he2018adversarial,  tang2019adversarial, yuan2019adversarial} for enhancing model robustness typically aim to preserve the overall accuracy metric against input perturbations. In other words, they stabilize the rank of one specific item (typically, the ground-truth next item) in a rank list rather than all the items in the list~\cite{yue2022defending, wu2021fight, anelli2021study, tan2023towards}, which can result in unstable rank lists after perturbations besides the position of that one item.

Our goal is to devise a stable recommender model that generates rank lists similar to the original rank lists, despite the presence of perturbations. We present a fine-tuning method for sequential recommender systems that maximizes model stability while preserving prediction performance, named \textbf{\method} (\underline{FINE}-tuning for \underline{ST}able Recommendations). 
To generate consistent rank lists with and without perturbations, \method requires \emph{reference} rank lists that can be used during the fine-tuning.
\method obtains reference rank lists for all training instances from a given pre-trained recommendation model.
\method can employ \textbf{any pre-trained recommendation model} to generate reference rank lists for fine-tuning, as long as the model's accuracy is comparable to the state-of-the-art.
Then, \method simulates a perturbation scenario by randomly sampling and perturbing a small number of interactions (e.g., 0.1\%) in each training epoch. 
After that, \method incorporates a \emph{novel regularization function} to encourage the rank lists generated by the model (being fine-tuned with the perturbed data) to be the same as the reference rank lists. 
This regularization function is optimized along with the next-item prediction objective on the top-$K$ items.
This fine-tuned model is used during test time as-is, regardless of various perturbations during testing.

\method is \emph{model-agnostic}, meaning that it can stabilize the recommendations for \emph{any} existing sequential recommender system against perturbations. Moreover, it is \emph{a fine-tuning method}, meaning that it can be applied to any \textit{pre-trained and even deployed} recommender system, while preserving model prediction accuracy. 
\method also \emph{empirically preserves recommendation performance} due to joint training of the next-item prediction objective and the rank-preserving objective.

The main contributions of this work include:
\begin{itemize}
\item 
\method is the \textbf{first} fine-tuning method that enhances the stability of \textbf{any} sequential recommenders against interaction-level perturbations, while maintaining or improving the prediction accuracy.  

\item
\method can improve the model stability against various types of perturbations via simulating perturbations during the fine-tuning. Its ranking-preserving regularization enables a model to preserve the ranking of items even in the presence of perturbations. \textbf{Our perturbation simulation and top-K-based self-distillation} are both unique compared to existing works.

\item
We validate both stability and accuracy of \method by comparing it with 5 fine-tuning mechanisms, on three real-world datasets, against diverse perturbation methods. Our results show that \method can considerably increase the stability of recommender systems without compromising the accuracy of model predictions, and that \method also \emph{empirically preserves recommendation performance} due to joint training of the next-item prediction objective and the rank-preserving objective. 

\end{itemize}

	\section{Related Work}
	\label{sec:related_work}
	\paragraph{Model Stability and Multiplicity in Machine Learning}
\label{sec:related_work_perturbation}

Our work is related to a stream of work in machine learning that highlights the sensitivity of model predictions to changes in the training data can produce significant changes in the output of a model~\cite{marx2020predictive, black2021leave, oh2022robustness}. This includes work on
predictive multiplicity~\cite{marx2020predictive,watsondaniels2022probabilistic} and underspecification~\cite{d2020underspecification}, which
show that supervised learning datasets can admit multiple prediction models that perform almost equally well yet assign different predictions on each test instance. A different stream of work highlights a similar degree of sensitivity that arise in settings where we update a model that is deployed by re-training it with a more recent dataset (i.e., ``predictive churn'')~\cite{milani2016launch}, or by removing a single instance from the training data~\cite{black2021leave}. Several methods have been developed to reduce this sensitivity as it leads to unexpected and harmful consequences in downstream applications and user experience~\cite{milani2016launch, jiang2021churn, hidey2022reducing}. Algorithms such as model regularization, distillation, and careful retraining~\cite{milani2016launch, jiang2021churn, hidey2022reducing} have been conducted to reduce prediction churn and stabilize model predictions. Our proposed method \method also utilizes a similar regularization technique to stabilize the recommender.

\paragraph{Adversarial Machine Learning}
\label{sec:related_work_adversarial_ML}
Adversarial training has been widely used in computer vision and natural language processing (NLP)~\cite{FGSM, morris2020textattack, liu2019mt-dnn} areas to enhance the robustness of deep learning models.
Many adversarial training methods use min-max optimization, which minimizes the maximal adversarial loss (i.e., worst-case scenario) computed with adversarial examples~\cite{wang2021adversarial}.
In computer vision, adversarial examples are generated by the Fast Gradient Sign Method~\cite{FGSM}, Projected Gradient Descent~\cite{athalye2018obfuscated}, or GANs~\cite{samangouei2018defense}, which can change the classification results. 
In the NLP area, adversarial examples are created in various ways, such as replacing characters or words in the input text and applying noise to input token embeddings~\cite{morris2020textattack, liu2019mt-dnn}. 
However, these models cannot be directly applied to recommender systems as they do not work on sequential interaction data or do not generate rank lists of items.

\paragraph{Robust Recommender Systems}
\label{sec:related_work_robust_recsys}
A large body of work on recommenders have primarily focused on improving accuracy; recently, there has been a surge of interest in addressing new emerging issues~\cite{ge2022survey, wang2022trustworthy} such as fairness~\cite{ekstrand2022fairness,wang2022survey}, diversity~\cite{castells2022novelty,SaMMLG22}, and robustness~\cite{Zhang2020PracticalDP, PoisonRec, taamr, wu2021triple, oh2022robustness}.
The majority of existing training or fine-tuning methods~\cite{wu2021fight, he2018adversarial, tang2019adversarial, park2019adversarial, yuan2019adversarial, du2018enhancing, anelli2021study, anelli2021formal, MSAP} for robust recommender systems are designed to provide accurate next-item predictions in the presence of input perturbations. 
However, as shown in \cref{tab:comparators}, most of these methods~\cite{wu2021fight, tang2019adversarial, tan2023towards, yue2022defending, park2019adversarial, yuan2019adversarial} have limitations in enhancing the ranking stability of sequential recommenders against input perturbations, as they are not optimized to preserve entire rank lists (but rather focus on ground-truth next-items)~\cite{wu2021fight, he2018adversarial, tan2023towards, yuan2019adversarial} or cannot be applied to sequential settings~\cite{park2019adversarial} (which can predict users' interests based on sequences of their recent interactions).  
A few of them~\cite{anelli2021study, tang2019adversarial} also require additional input like images. While a few ranking-distillation methods~\cite{yue2021black,tang2018ranking} can be adapted to our setting, they are unsuitable for preserving rank lists against input perturbations, as they may sacrifice the next-item prediction accuracy to achieve their goal.

\begin{table}[t]
	\centering
	\caption{Overview of existing methods to build robust recommender systems.}
	\resizebox{\textwidth}{!}{
 \begin{tabular}{lccccccc}
		\textbf{Functionality} & \textbf{\method}
        & APT~\cite{wu2021fight}
		&  AMR~\cite{tang2019adversarial}
		&  
  AdvT~\cite{yue2022defending}
		&
  CAT~\cite{tan2023towards} 
		&  AdvIR~\cite{park2019adversarial} & 
  ACAE~\cite{yuan2019adversarial}  
		 \\
		\midrule
  Sequential Recommender & \textbf{\checkmark} &  &  & \checkmark &   \checkmark & &  \checkmark \\
  No Need to Retrain the Model & \textbf{\checkmark} &  \checkmark & \checkmark &  &  & & \checkmark \\
	Preserve All Users' Rank Lists & \textbf{\checkmark}  &  &  &  &  &  &  \\
  Ensure Model Robustness & \textbf{\checkmark} &  \checkmark &   \checkmark & \checkmark &  \checkmark &  \checkmark & \checkmark \\
	\end{tabular}
 }
	\label{tab:comparators}
\end{table}
	
	\section{Preliminaries}
	\label{sec:prelim:cascading}
	\subsection{Sequential Recommendations}
\label{sec:prelim:sequential_recsys}
We focus on sequential recommender models in this work, which are trained to accurately predict the next item of a user based on their previous interactions.
Formally, we have a set of users $\mathcal{U}$ and items $\mathcal{I}$. For a user $u$, their interactions are represented as a sequence of items (sorted by timestamps) denoted as $S^{u} = \{S_{1}^{u}, \ldots S_{m^u}^{u}\}$, and the corresponding timestamps as $T^{u} = \{T_{1}^{u}, \ldots T_{m^u}^{u}\}$. Here, $S_{t}^{u} \in \mathcal{I}$, and $m^u$ represents the total number of interactions for a user $u$. 
We train the sequential recommender with the following loss function to predict the next item $S_{t+1}^{u}$ accurately for each user $u$, given an item sequence $\{S_{1}^{u}, \ldots , S_{t}^{u}\}, \forall t \in [1, m^u-1]$.
\begin{equation}
\mathcal{L} = \sum_{u \in \mathcal{U}}\sum_{t=1}^{m^u-1} CE(\mathbf{1}^{S_{t+1}^{u}}, \Theta(\{S_{1}^{u}, \ldots , S_{t}^{u}\})).
\label{eq:loss_origin}
\end{equation}
Note that $\mathbf{1}^{i} \in \mathbb{R}^{|\mathcal{I}|}$ is a one-hot vector where the $i^{th}$ value is 1, $CE$ represents the Cross-Entropy function, 
and $\Theta (\{S_{1}^{u}, \ldots , S_{t}^{u}\})  \in \mathbb{R}^{|\mathcal{I}|}$ is the next-item prediction vector generated by the model for a user $u$, given their historical item sequence $\{S_{1}^{u}, \ldots , S_{t}^{u}\}$. 
Given the sequential data, we define an instance ($X_n$) as a pair consisting of an observed item sequence and the ground-truth next-item, i.e., $(\{S_{1}^{u}, \ldots , S_{t}^{u}\}, S_{t+1}^{u})$.

\subsection{Measuring Model Stability} 
\label{sec:prelim:model_stability}
Recent work~\cite{oh2022robustness} has shown that existing sequential recommenders generate unstable predictions when subjected to input data perturbations.
Specifically, consider two scenarios. 
First, a recommender model $\Theta$ is trained on the original training data and generates recommendation lists $R^{X_n}_{\Theta}$ for all test instances $X_n$ in the test data $X_{\mathit{test}}$.
Second, another model $\Theta'$, which shares the same initial parameters as $\Theta$, is trained on perturbed training data and produces rank lists $R^{X_n}_{\Theta'}$ for all $X_n \in X_{\mathit{test}}$.
If the original model $\Theta$ is robust against input perturbation, then $R^{X_n}_\Theta$ and $R^{X_n}_{\Theta'}$ should be highly similar for all $X_n \in X_{\mathit{test}}$.
However, existing work~\cite{oh2022robustness} has shown that even if there are minor changes to the training data, then $R^{X_n}_{\Theta'}$ is drastically different from $R^{X_n}_\Theta$ for all $X_n \in X_{\mathit{test}}$.
To quantify the stability of the model $\Theta$ against input perturbations, we use the Rank List Stability (RLS)~\cite{oh2022robustness} metric:
\begin{equation}
RLS = \frac{1}{|X_{\mathit{test}}|}\sum_{\forall X_n \in X_{\mathit{test}}} \mathit{similarity}(R^{X_n}_\Theta, R^{X_n}_{\Theta'}),    
\label{eq:RLS}
\end{equation}
where $\mathit{similarity}(A,B)$ denotes a similarity function between two rank lists $A$ and $B$. Following \cite{oh2022robustness},  we use Rank-biased Overlap (RBO)~\cite{RBO} and Top-$K$ Jaccard Similarity~\cite{jaccard1912distribution} as similarity functions. 

\noindent (1) {\textbf{Rank-biased Overlap (RBO)}}: 
RBO~\cite{RBO} measures the similarity of orderings between two rank lists. A higher RBO indicates two rank lists are similar, and RBO values lie between 0 and 1. RBO gives higher importance to the similarity in the top part of the rank lists than the bottom part, making it our primary metric and preferable compared to other metrics like Kendall's Tau~\cite{kendall1948rank}.
RBO of two rank lists $A$ and $B$ with $|\mathcal{I}|$ items is defined as follows.
$$ \mathit{RBO} (A,B) = (1-p) \sum_{d=1}^{|\mathcal{I}|}{p^{d-1}\frac{|A[1:d] \cap B[1:d]|}{d}},$$
where $p$ is a hyperparameter (recommended value: 0.9). 
    
\noindent (2) \textbf{Top-$K$ Jaccard similarity}: 
    Jaccard similarity ($\mathit{Jaccard}(A,B) = \frac{|A \cap B|}{|A \cup B|}$) is used to calculate the ratio of common top-$K$ items between two rank lists, without considering the item ordering. The score ranges from 0 to 1, and a higher score indicates the top-$K$ items in two rank lists are similar (indicating higher model stability).
    Regarding $K$, we use $K = 10$ as it is common practice~\cite{JODIE, hansen2020contextual}.
    
\subsection{Scope: Interaction-level Perturbations}
\label{sec:prelim:scope}
Similar to prior work~\cite{oh2022robustness}, we assume \textbf{interaction-level minor perturbations}. Interaction perturbations are the smallest perturbation compared to user or item perturbations. Minor perturbations indicate that only a small number of interactions (e.g., 0.1\%) can be perturbed in the training data. 
Naturally, larger perturbations will result in a greater decrease in stability.
Possible perturbations include injecting interactions, deleting interactions, replacing items of interactions with other items, or a mix of them. To find such interactions to perturb,  
we can employ various perturbation algorithms for recommender systems~\cite{Revisiting_RecSys, yue2021black, oh2022robustness, pruthi2020estimating, zhang2021data, wu2021triple}.

For instance, \casper~\cite{oh2022robustness} is the state-of-the-art interaction perturbation method for sequential recommenders.
\casper defines a \textit{cascading score} of a training instance $X_k$ as the number of training interactions that will be affected if $X_k$ is perturbed.
To compute the cascading score, \casper creates an interaction-to-interaction dependency graph (IDAG) based on the training data, which encodes the influence of one interaction on another. 
Given this IDAG, the cascading score of an instance $X_k$ is defined as the number of descendants of $X_k$ in the graph (i.e., 
all the nodes reachable from $X_k$ by following the outgoing edges in the IDAG). 
Among all training instances, perturbing an instance with the largest cascading score leads to the maximal changes in test-time recommendations after model retraining.

	\section{Problem Setup}
	\label{sec:proposed_method}
	\begin{figure*}[t!]
    \centering
    \includegraphics[width=1.0\linewidth]{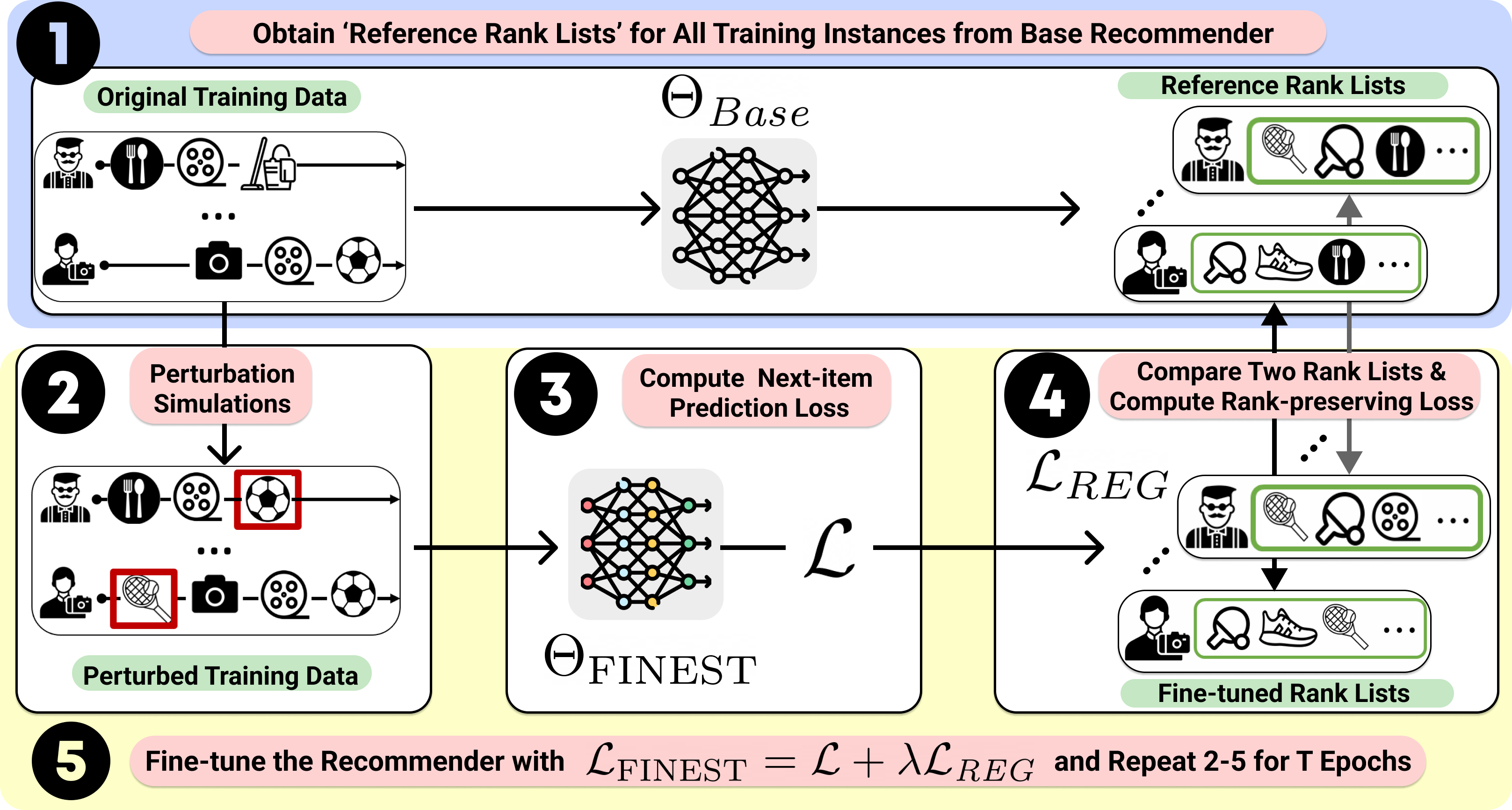}
    \caption{
 Overview of stabilizing a recommender model via fine-tuning with \method. First, we obtain reference recommendations for all training instances from a given recommendation model. Next, randomly sampled and perturbed data (changing every epoch) is to fine-tune the model. \method\ adds rank-preserving regularization to minimize differences between the reference and fine-tuned rank lists (generated under pseudo-perturbations). By simulating perturbations, \method can generate stable rank lists even in the presence of actual input perturbations. 
}
	\label{fig:main_plot}
\end{figure*}

\subsection{Goal} Let $\Theta$ and $\Theta'$ be recommendation models trained with the original and perturbed training data, respectively. Also, let $R^{X_n}_\Theta$ and $R^{X_n}_{\Theta'}$ be the rank lists generated for a test instance $X_n$ without and with perturbations, respectively. Then, our goal is to fine-tune $\Theta'$ with our proposed method \method, so that $R^{X_n}_{\Theta'}$ would be identical to $R^{X_n}_\Theta$ for all $X_n \in X_\mathit{test}$. Let $A[i]$ represent the $i^{th}$ item in a rank list $A$, and $|\mathcal{I}|$ be the number of items. Then, formally, our objective is to ensure that:  $R^{X_n}_{\Theta'}[i] = R^{X_n}_\Theta[i], \forall i$ from $1$ to $|\mathcal{I}|$ for all $X_n \in X_\mathit{test}$ after fine-tuning 
with \method.

\subsection{Assumptions}
\noindent (1)  The specific training interactions that are perturbed are not known during the fine-tuning. 
Thus, a fine-tuning model needs to be created that is robust regardless of the various interaction perturbations.

\noindent (2) As model designers aiming to increase the model stability, we naturally have access to all the training data and the recommendation model (e.g., model parameters).

\subsection{Measuring Training Effectiveness against Perturbations}
We first measure the stability of a base recommender model $\Theta_\mathit{Base}$, which is trained with the typical next-item prediction objective.
The stability of  ${\Theta_\mathit{Base}}$ is quantified by the RLS metrics using Equation~\eqref{eq:RLS}, where high RLS values indicate the model is stable. 
Next, we fine-tune ${\Theta_\mathit{Base}}$ with \method (we call this fine-tuned model $\Theta_\mathit{\method}$) and compute its stability using the RLS metrics. 
\method is successful if the stability of $\Theta_\mathit{\method}$ is higher than  ${\Theta_\mathit{Base}}$.

\section{Proposed Methodology}
\label{sec:method:overview}
We introduce a fine-tuning method called \method to enhance the stability of sequential recommender systems against input perturbations. 
\method simulates perturbation scenarios in the training data and aims to maximize the model's stability against such emulated perturbations with a rank-aware regularization function. \cref{fig:main_plot} summarizes the main steps of \method. 
In Step 1, the base recommender $\Theta_\mathit{Base}$ is used to generate ranked item lists for all training instances $X_n \in X_\mathit{train}$: $R^{X_n}_{\Theta_\mathit{Base}}$.
These lists will serve as reference rank lists for fine-tuning the model. 
Next, the base recommender $\Theta_\mathit{Base}$  is fine-tuned with \method for $T$ epochs (Steps 2--5), with $T$ being a hyperparameter. 

\begin{algorithm} [t!]
	\small
	\caption{\textbf{\method}: \textbf{FINE}-tuning for model \textbf{ST}ability} \label{alg:main}
	\SetKwInOut{Input}{Input}
	\SetKwInOut{Output}{Output}
	\Input{ 
		A base recommender $\Theta_\mathit{Base}$, training data $X_\mathit{train}$, sampling ratio $R$, number of sampled items $K$, number of fine-tuning epochs $T$, regularization constants $\lambda, \lambda_1, \lambda_2$.\\
	}
	\Output{
    A fine-tuned recommender $\Theta_{\method}$\\
	} 
	    \Comment{\textbf{Step 1. Generate Reference Rank Lists }}{
	Generate reference rank lists $R^{X_n}_{\Theta_\mathit{Base}}, \forall X_n \in X_\mathit{train}$ using a given recommendation model $\Theta_\mathit{Base}$ }
 \\
  $\Theta_\mathit{\method} \longleftarrow \Theta_\mathit{Base}$
    \\
	\For{Fine-tuning epoch $\in [1, \ldots, T]$}{
	    \Comment{\textbf{Step 2. Training data is pseudo-perturbed }}{
	        Perform random sampling of interactions with the ratio $R$\\
	        Perturb interactions by deletion, replacement, or insertion with equal probability; $X_\mathit{pert} \longleftarrow$ the set of perturbed interactions\\
         }
	    \Comment{\textbf{Step 3. Perform Next-item Prediction}}{
	        Calculate the loss $\mathcal{L}$ using \cref{eq:loss_origin} with perturbed training data \\
	    }
	    \Comment{\textbf{Step 4. Rank-preserving Regularization}}{
	        Compute recommendation prediction scores of top-$2K$ items using $\Theta_{\method}$ for all training instances in $X_{n} \in X_\mathit{train} \backslash X_\mathit{pert}$ \\
	        Compute the regularization loss $\mathcal{L}_\mathit{REG}$ using \cref{eq:topK_regularization_all}\\
	    }
	    \Comment{\textbf{Step 5. Fine-tune to Stabilize Rank Lists}}{
	    Update the model parameters $\Theta_{\method}$ using \cref{eq:loss_adversarial}\\
	    }
	}
    Return the fine-tuned recommendation model $\Theta_{\method}$ \\
\end{algorithm}

\subsection{Perturbation Simulations}
\label{sec:method:sampling}
Applying random perturbations on training data has contributed to enhancing model stability against input data perturbations in computer vision~\cite{rosenfeld2020certified, gong2021maxup, levine2020robustness} and NLP~\cite{Swenor_2022}. Taking inspiration from this, 
\method simulates a pseudo-perturbation by randomly sampling training interactions (with a sampling ratio $R$) and perturbing them every epoch. Perturbations include one of three actions with equal probability: the interaction can be \textit{deleted}, the interaction's item can be \textit{replaced} with another, or a new interaction can be \textit{inserted} before it.
Re-sampling in every epoch ensures that the recommender model sees many variations of the input data and is able to learn to make accurate and stable predictions regardless of a specific perturbation.

\cref{fig:main_plot} depicts an example of perturbing training data by insertion. In an insertion perturbation of an interaction $(u,i,t)$, we inject the least popular item into $u$'s sequence with a timestamp right before $t$ (e.g., $t-1$). Similarly, in an item replacement perturbation, the target item of the interaction is replaced with the least popular item in the dataset. In both cases, the least popular item is selected as it leads to the lowest RLS metrics of the recommender model compared to other items~\cite{oh2022robustness}.

\subsection{Next-item Prediction on Perturbed Simulations} 
The perturbed data is used to fine-tune $\Theta_\mathit{Base}$. Specifically, the parameters of $\Theta_\mathit{Base}$ are used to initialize $\Theta_\mathit{\method}$. It is then fine-tuned for $T$ epochs on the next-item prediction loss using the perturbed data as input (see \cref{eq:loss_origin}). This loss will be added with the rank-preserving regularization loss in the next step. 

\subsection{Rank-preserving  Regularization}
\label{sec:method:regularization}
Due to the pseudo-perturbations, the rank lists generated by a current model (being fine-tuned) may differ from a reference rank list. 
Rank-preserving regularization aims to ensure that two rank lists are identical after fine-tuning, meaning the rank lists do not change as a result of the perturbation.

Ideally, we need to preserve the ranking of all items in the reference rank lists to ensure full stability. In practice, maintaining the ranks of all items in all the rank lists is computationally prohibitive as there are millions of items in most recommendation tasks (e.g., in e-commerce~\cite{smith2017two}). 
Therefore, to make our method \method scalable, we only aim to preserve the rank of the top-$K$ items -- that are displayed to users typically -- from each reference rank list.
Formally, for each training instance $X_n \in X_\mathit{train}$ and its reference rank list $R^{X_n}_{\Theta_\mathit{B}}$, we sample the top-$2K$ items $R^{X_n}_{\Theta_\mathit{B}}[1:2K]$, where
$\Theta_B$ and $\Theta_F$ are shorthands for $\Theta_{Base}$ and $\Theta_{\method}$, respectively.
Let $\{i_1, \ldots, i_{2K}\}$ denote the top-$2K$ items in the reference rank list $R^{X_n}_{\Theta_B}[1:2K]$.
We underscore that \textbf{$\{i_1, \ldots, i_{2K}\}$ are obtained from the reference rank list $R^{X_n}_{\Theta_B}$}, not the fine-tuned rank list  $R^{X_n}_{\Theta_F}$.
Using the sampled items, \method creates a novel rank-aware regularization loss
$\mathcal{L}_\mathit{REG}$.

\begin{figure}[t!]
    \includegraphics[width=0.6\linewidth]{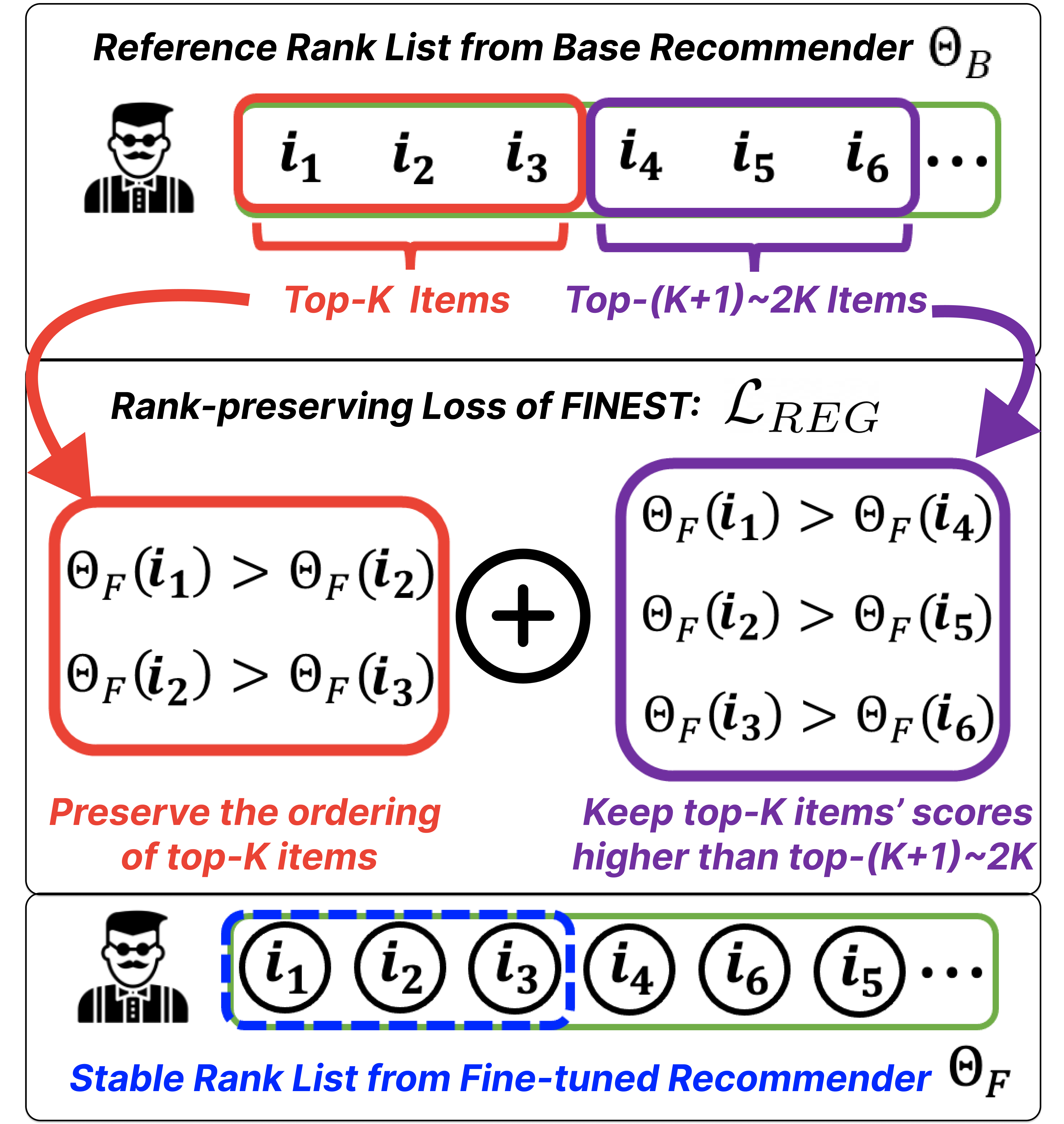}
    \caption{
    \method uses a rank-preserving regularization term (\cref{eq:topK_regularization_single}) to penalize differences in ordering and prediction scores of the top-$K$ items with respect to a reference rank list. With the regularizer, the recommender can generate a similar top-$K$ recommendation to the reference one under perturbations. 
    $\Theta_{B}$ and $\Theta_{F}$ indicate   $\Theta_\mathit{Base}$ and $\Theta_{\method}$, respectively.
    }
	\label{fig:rank_preserving_loss}
\end{figure}

\begin{equation}
\begin{aligned}
\mathcal{L}_\mathit{REG}(X_n) = \overbrace{ \sum_{k=1}^{K-1} \max(\Theta_F(i_{k+1}) -\Theta_F(i_{k})+\lambda_1,0) }^{\mathclap{\text{penalize violations in relate order of reference top-$K$ items}}}
\\+\underbrace{\sum_{k=1}^{K} \max(\Theta_F(i_{k+K}) - \Theta_F(i_{k})+\lambda_2,0)}_{\mathclap{\text{penalize if higher prediction scores are assigned to reference top-$(K+1)$ to top-$2K$ items than reference top-$K$ items}}},
\end{aligned}
\label{eq:topK_regularization_single}
\end{equation}

\noindent 
Here $\lambda_1$ and $\lambda_2$ are margin values (hyperparameters; user-specified). 
%
The first loss term penalizes if $\Theta_F$ gives a higher prediction score to $i_{k+1}$ compared to $i_k$, which ensures that the relative ordering of top-$K$ items in $R^{X_n}_{\Theta_B}$ is also maintained in $R^{X_n}_{\Theta_F}$. 
The second loss term penalizes if $\Theta_F$ gives higher prediction scores to ``competitive'' items (i.e., $\{i_{K+1}, \ldots, i_{2K}\}$) than to the desired top-$K$ items $\{i_1, \ldots, i_{K}\}$, which places the top-$K$ items from the reference rank list at the top part of the fine-tuned rank list. 
Together these terms ensure that the relative positions and ordering of the top-$K$ sampled items are the same in both rank lists as shown in \cref{fig:rank_preserving_loss}.

The total regularization loss over all non-perturbed training instances is defined as follows:
\begin{equation}
\begin{aligned}
\mathcal{L}_\mathit{REG} = \sum_{\forall X_n \in X_\mathit{train} \backslash X_\mathit{pert}} \mathcal{L}_\mathit{REG}(X_n),
\end{aligned}
\label{eq:topK_regularization_all}
\end{equation}

\noindent where $X_\mathit{pert}$ is the set of perturbed instances in the current epoch.
$\mathcal{L}_\mathit{REG}$ is computed only for all non-perturbed instances $X_\mathit{train} \backslash X_\mathit{pert}$ of the current epoch. 
Perturbed instances are excluded because reference rank lists can be unavailable for perturbed instances.

It is important to \emph{highlight the difference between the distillation loss~\cite{yue2021black} and the proposed rank-preserving loss}. \citet{yue2021black} only choose randomly sampled negative items as ``competitors'' in the second loss term of \cref{eq:topK_regularization_single}. 
However, the random selection scheme does not help in preserving top-$K$ ranking if the chosen negative samples are low-ranked in the reference rank lists.

\subsection{Total Loss}

Overall, the total loss of \method simultaneously optimizes for the next-item prediction performance in \cref{eq:loss_origin} and the rank-preserving regularization performance in \cref{eq:topK_regularization_all} as follows:
\begin{equation}
\mathcal{L}_\mathit{\method} = \mathcal{L} + \lambda \mathcal{L}_\mathit{REG},
\label{eq:loss_adversarial}
\end{equation}
where $\lambda$ is a hyperparameter (user-specified) that controls the regularization strength .
It is essential to optimize both objectives together. If the next-item prediction loss $\mathcal{L}$ is not included, then the model may sacrifice next-item prediction performance in favor of the stability objective. This is undesirable as it will reduce the utility of the resulting model.

	\section{Experiments}
	\label{sec:experiment}
	
\begin{table}[t!]
	\centering
	\caption{Recommendation datasets used for experiments.
	}
	\centering
	\begin{tabular}{ l  r r r r l}
		\textbf{Name} & \textbf{Users} & \textbf{Items} & \textbf{Interactions} & \textbf{Descriptions}\\
		\midrule
		LastFM & 980 & 1,000 &  1,293,103 & Music playing history  \\
		Foursquare & 2,106 & 5,597  & 192,602 & Point-of-Interest check-in\\
		Reddit & 4,675 & 953 & 134,489 & Subreddit posting history\\
	\bottomrule
	\end{tabular}	
	\label{tab:dataset}
\end{table}

\begin{table*}[t!]
\caption{Effectiveness of various fine-tuning methods for recommendation models on top-2 largest datasets. \method is the best fine-tuning method as per enhancing model stability (measured by RLS metrics)
against random and \casper~\cite{oh2022robustness} deletion perturbations, with statistical significance (p-values < 0.05) in all cases. \method also performs the best for other types of perturbations and datasets.}
\begin{subtable}[h]{1.0\linewidth}
\footnotesize
\caption{LastFM Dataset (Music Recommendation; 1.3 Million Interactions)}
\label{tab:defense_lastfm}
\begin{tabular}{|c|cccccc|cccccc|}
\hline
\textbf{Perturbations} &
  \multicolumn{6}{c|}{\textbf{Random Deletion Perturbations}} &
  \multicolumn{6}{c|}{\textbf{\casper~\cite{oh2022robustness} Deletion Perturbations}} \\ \hline
\textbf{Recommenders} &
  \multicolumn{2}{c|}{\textbf{\tisasrec~\cite{Tisasrec}}} &
  \multicolumn{2}{c|}{\textbf{\bert~\cite{sun2019bert4rec}}} &
  \multicolumn{2}{c|}{\textbf{\lstm~\cite{hochreiter1997long}}} &
  \multicolumn{2}{c|}{\textbf{\tisasrec~\cite{Tisasrec}}} &
  \multicolumn{2}{c|}{\textbf{\bert~\cite{sun2019bert4rec}}} &
  \multicolumn{2}{c|}{\textbf{\lstm~\cite{hochreiter1997long}}} \\ \hline
\textbf{RLS Metrics} &
  \multicolumn{1}{c|}{\textbf{RBO}} &
  \multicolumn{1}{c|}{\textbf{Jaccard}} &
  \multicolumn{1}{c|}{\textbf{RBO}} &
  \multicolumn{1}{c|}{\textbf{Jaccard}} &
  \multicolumn{1}{c|}{\textbf{RBO}} &
  \multicolumn{1}{c|}{\textbf{Jaccard}} &
  \multicolumn{1}{c|}{\textbf{RBO}} &
  \multicolumn{1}{c|}{\textbf{Jaccard}} &
  \multicolumn{1}{c|}{\textbf{RBO}} &
  \multicolumn{1}{c|}{\textbf{Jaccard}} &
  \multicolumn{1}{c|}{\textbf{RBO}} &
  \multicolumn{1}{c|}{\textbf{Jaccard}} \\ \hline
\textbf{Original} &
  \multicolumn{1}{c|}{\cellcolor{red!25} 0.753} &
  \multicolumn{1}{c|}{\cellcolor{red!25} 0.275} &
  \multicolumn{1}{c|}{\cellcolor{red!25} 0.754} &
  \multicolumn{1}{c|}{\cellcolor{red!25} 0.316} &
  \multicolumn{1}{c|}{\cellcolor{red!25} 0.769} &
  \multicolumn{1}{c|}{\cellcolor{red!25} 0.269} 
  &
  \multicolumn{1}{c|}{\cellcolor{red!25} 0.694} &
  \multicolumn{1}{c|}{\cellcolor{red!25} 0.200} &
  \multicolumn{1}{c|}{\cellcolor{red!25} 0.754} &
  \multicolumn{1}{c|}{\cellcolor{red!25} 0.316} &
  \multicolumn{1}{c|}{\cellcolor{red!25} 0.700} &
  \multicolumn{1}{c|}{\cellcolor{red!25} 0.172} 
   \\ \hline
\textbf{Random} &
  \multicolumn{1}{c|}{0.762} &
  \multicolumn{1}{c|}{0.295} &
  \multicolumn{1}{c|}{0.776} &
  \multicolumn{1}{c|}{0.373} &
  \multicolumn{1}{c|}{0.787} &
  \multicolumn{1}{c|}{0.300} 
   &
  \multicolumn{1}{c|}{0.702} &
  \multicolumn{1}{c|}{0.215} &
  \multicolumn{1}{c|}{0.773} &
  \multicolumn{1}{c|}{0.366} &
  \multicolumn{1}{c|}{0.699} &
  \multicolumn{1}{c|}{0.167} 
   \\ 
\textbf{Earliest-Random} &
  \multicolumn{1}{c|}{0.760} &
  \multicolumn{1}{c|}{0.290} &
  \multicolumn{1}{c|}{0.776} &
  \multicolumn{1}{c|}{0.364} &
  \multicolumn{1}{c|}{0.780} &
  \multicolumn{1}{c|}{0.292} 
   &
  \multicolumn{1}{c|}{0.702} &
  \multicolumn{1}{c|}{0.212} &
  \multicolumn{1}{c|}{0.774} &
  \multicolumn{1}{c|}{0.364} &
  \multicolumn{1}{c|}{0.700} &
  \multicolumn{1}{c|}{0.168} 
   \\
\textbf{Latest-Random} &
  \multicolumn{1}{c|}{0.763} &
  \multicolumn{1}{c|}{0.302} &
  \multicolumn{1}{c|}{0.784} &
  \multicolumn{1}{c|}{0.380} &
  \multicolumn{1}{c|}{0.774} &
  \multicolumn{1}{c|}{0.280} 
   &
  \multicolumn{1}{c|}{0.703} &
  \multicolumn{1}{c|}{0.220} &
  \multicolumn{1}{c|}{0.777} &
  \multicolumn{1}{c|}{0.367} &
  \multicolumn{1}{c|}{0.699} &
  \multicolumn{1}{c|}{0.166} 
   \\ 
\textbf{APT~\cite{wu2021fight}} &
  \multicolumn{1}{c|}{0.764} &
  \multicolumn{1}{c|}{0.297} &
  \multicolumn{1}{c|}{0.777} &
  \multicolumn{1}{c|}{0.368} &
  \multicolumn{1}{c|}{0.779} &
  \multicolumn{1}{c|}{0.290} 
   &
  \multicolumn{1}{c|}{0.701} &
  \multicolumn{1}{c|}{0.212} &
  \multicolumn{1}{c|}{0.775} &
  \multicolumn{1}{c|}{0.363} &
  \multicolumn{1}{c|}{0.699} &
  \multicolumn{1}{c|}{0.168} 
   \\
\textbf{ACAE~\cite{yuan2019adversarial}} &
  \multicolumn{1}{c|}{0.763} &
  \multicolumn{1}{c|}{0.294} &
  \multicolumn{1}{c|}{0.770} &
  \multicolumn{1}{c|}{0.352} &
  \multicolumn{1}{c|}{0.779} &
  \multicolumn{1}{c|}{0.286} 
   &
  \multicolumn{1}{c|}{0.699} &
  \multicolumn{1}{c|}{0.210} &
  \multicolumn{1}{c|}{0.773} &
  \multicolumn{1}{c|}{0.365} &
  \multicolumn{1}{c|}{0.700} &
  \multicolumn{1}{c|}{0.169} 
   \\ \hline
\textbf{\method} &
  \multicolumn{1}{c|}{\cellcolor{blue!25} \textbf{0.921}} &
  \multicolumn{1}{c|}{\cellcolor{blue!25} \textbf{0.659}} &
  \multicolumn{1}{c|}{\cellcolor{blue!25} \textbf{0.835}} &
  \multicolumn{1}{c|}{\cellcolor{blue!25} \textbf{0.482}} &
  \multicolumn{1}{c|}{\cellcolor{blue!25} \textbf{0.904}} &
  \multicolumn{1}{c|}{\cellcolor{blue!25} \textbf{0.590}}
   &
  \multicolumn{1}{c|}{\cellcolor{blue!25} \textbf{0.873}} &
  \multicolumn{1}{c|}{\cellcolor{blue!25} \textbf{0.519}} &
  \multicolumn{1}{c|}{\cellcolor{blue!25} \textbf{0.832}} &
  \multicolumn{1}{c|}{\cellcolor{blue!25} \textbf{0.476}} &
  \multicolumn{1}{c|}{\cellcolor{blue!25} \textbf{0.787}} &
  \multicolumn{1}{c|}{\cellcolor{blue!25} \textbf{0.335}}
   \\ \hline
   \textbf{\% Improvements} &
  \multicolumn{1}{c|}{\cellcolor{blue!25} \textbf{+21 \%}} &
  \multicolumn{1}{c|}{\cellcolor{blue!25} \textbf{+119\%}} &
  \multicolumn{1}{c|}{\cellcolor{blue!25} \textbf{+6.5\%}} &
  \multicolumn{1}{c|}{\cellcolor{blue!25} \textbf{+27\%}} &
  \multicolumn{1}{c|}{\cellcolor{blue!25} \textbf{+15\%}} &
  \multicolumn{1}{c|}{\cellcolor{blue!25} \textbf{+97\%}}
   &
  \multicolumn{1}{c|}{\cellcolor{blue!25} \textbf{+24\%}} &
  \multicolumn{1}{c|}{\cellcolor{blue!25} \textbf{+136\%}} &
  \multicolumn{1}{c|}{\cellcolor{blue!25} \textbf{+7.5\%}} &
  \multicolumn{1}{c|}{\cellcolor{blue!25} \textbf{+30\%}} &
  \multicolumn{1}{c|}{\cellcolor{blue!25} \textbf{+12\%}} &
  \multicolumn{1}{c|}{\cellcolor{blue!25} \textbf{+95\%}}
   \\ \hline
\end{tabular}
\end{subtable}
\begin{subtable}[h]{1.0\linewidth}
\footnotesize
\caption{Foursquare Dataset (POI Recommendation; 0.2 Million Interactions)}
\label{tab:defense_foursquare}
\begin{tabular}{|c|cccccc|cccccc|}
\hline
\textbf{Perturbations} &
  \multicolumn{6}{c|}{\textbf{Random Deletion Perturbations}} &
  \multicolumn{6}{c|}{\textbf{\casper~\cite{oh2022robustness} Deletion Perturbations}} \\ \hline
\textbf{Recommenders} &
  \multicolumn{2}{c|}{\textbf{\tisasrec~\cite{Tisasrec}}} &
  \multicolumn{2}{c|}{\textbf{\bert~\cite{sun2019bert4rec}}} &
  \multicolumn{2}{c|}{\textbf{\lstm~\cite{hochreiter1997long}}} &
  \multicolumn{2}{c|}{\textbf{\tisasrec~\cite{Tisasrec}}} &
  \multicolumn{2}{c|}{\textbf{\bert~\cite{sun2019bert4rec}}} &
  \multicolumn{2}{c|}{\textbf{\lstm~\cite{hochreiter1997long}}} \\ \hline
\textbf{RLS Metrics} &
  \multicolumn{1}{c|}{\textbf{RBO}} &
  \multicolumn{1}{c|}{\textbf{Jaccard}} &
  \multicolumn{1}{c|}{\textbf{RBO}} &
  \multicolumn{1}{c|}{\textbf{Jaccard}} &
  \multicolumn{1}{c|}{\textbf{RBO}} &
  \multicolumn{1}{c|}{\textbf{Jaccard}} &
  \multicolumn{1}{c|}{\textbf{RBO}} &
  \multicolumn{1}{c|}{\textbf{Jaccard}} &
  \multicolumn{1}{c|}{\textbf{RBO}} &
  \multicolumn{1}{c|}{\textbf{Jaccard}} &
  \multicolumn{1}{c|}{\textbf{RBO}} &
  \multicolumn{1}{c|}{\textbf{Jaccard}} \\ \hline
\textbf{Original} &
  \multicolumn{1}{c|}{\cellcolor{red!25} 0.768} &
  \multicolumn{1}{c|}{\cellcolor{red!25} 0.273} &
  \multicolumn{1}{c|}{\cellcolor{red!25} 0.795} &
  \multicolumn{1}{c|}{\cellcolor{red!25} 0.354} &
  \multicolumn{1}{c|}{\cellcolor{red!25} 0.710} &
  \multicolumn{1}{c|}{\cellcolor{red!25} 0.168} 
  &
  \multicolumn{1}{c|}{\cellcolor{red!25} 0.779} &
  \multicolumn{1}{c|}{\cellcolor{red!25} 0.284} &
  \multicolumn{1}{c|}{\cellcolor{red!25} 0.796} &
  \multicolumn{1}{c|}{\cellcolor{red!25} 0.357} &
  \multicolumn{1}{c|}{\cellcolor{red!25} 0.646} &
  \multicolumn{1}{c|}{\cellcolor{red!25} 0.114}
   \\ \hline
\textbf{Random} &
  \multicolumn{1}{c|}{0.763} &
  \multicolumn{1}{c|}{0.262} &
  \multicolumn{1}{c|}{0.819} &
  \multicolumn{1}{c|}{0.428} &
  \multicolumn{1}{c|}{0.708} &
  \multicolumn{1}{c|}{0.171}
   &
  \multicolumn{1}{c|}{0.769} &
  \multicolumn{1}{c|}{0.266} &
  \multicolumn{1}{c|}{0.815} &
  \multicolumn{1}{c|}{0.415} &
  \multicolumn{1}{c|}{0.647} &
  \multicolumn{1}{c|}{0.118}
   \\ 
\textbf{Earliest-Random} &
  \multicolumn{1}{c|}{0.764} &
  \multicolumn{1}{c|}{0.257} &
  \multicolumn{1}{c|}{0.811} &
  \multicolumn{1}{c|}{0.404} &
  \multicolumn{1}{c|}{0.715} &
  \multicolumn{1}{c|}{0.177}
   &
  \multicolumn{1}{c|}{0.774} &
  \multicolumn{1}{c|}{0.266} &
  \multicolumn{1}{c|}{0.815} &
  \multicolumn{1}{c|}{0.414} &
  \multicolumn{1}{c|}{0.648} &
  \multicolumn{1}{c|}{0.118}
   \\
\textbf{Latest-Random} &
  \multicolumn{1}{c|}{0.758} &
  \multicolumn{1}{c|}{0.255} &
  \multicolumn{1}{c|}{0.815} &
  \multicolumn{1}{c|}{0.415} &
  \multicolumn{1}{c|}{0.713} &
  \multicolumn{1}{c|}{0.175} 
   &
  \multicolumn{1}{c|}{0.762} &
  \multicolumn{1}{c|}{0.263} &
  \multicolumn{1}{c|}{0.816} &
  \multicolumn{1}{c|}{0.417} &
  \multicolumn{1}{c|}{0.646} &
  \multicolumn{1}{c|}{0.116} 
   \\ 
\textbf{APT~\cite{wu2021fight}} &
  \multicolumn{1}{c|}{0.762} &
  \multicolumn{1}{c|}{0.297} &
  \multicolumn{1}{c|}{0.808} &
  \multicolumn{1}{c|}{0.392} &
  \multicolumn{1}{c|}{0.708} &
  \multicolumn{1}{c|}{0.169}
   &
  \multicolumn{1}{c|}{0.790} &
  \multicolumn{1}{c|}{0.325} &
  \multicolumn{1}{c|}{0.809} &
  \multicolumn{1}{c|}{0.400} &
  \multicolumn{1}{c|}{0.648} &
  \multicolumn{1}{c|}{0.117}
   \\
\textbf{ACAE~\cite{yuan2019adversarial}} &
  \multicolumn{1}{c|}{0.780} &
  \multicolumn{1}{c|}{0.292} &
  \multicolumn{1}{c|}{0.805} &
  \multicolumn{1}{c|}{0.383} &
  \multicolumn{1}{c|}{0.714} &
  \multicolumn{1}{c|}{0.169}
   &
  \multicolumn{1}{c|}{0.787} &
  \multicolumn{1}{c|}{0.303} &
  \multicolumn{1}{c|}{0.807} &
  \multicolumn{1}{c|}{0.388} &
  \multicolumn{1}{c|}{0.650} &
  \multicolumn{1}{c|}{0.119}
   \\ \hline
\textbf{\method} &
  \multicolumn{1}{c|}{\cellcolor{blue!25} \textbf{0.937}} &
  \multicolumn{1}{c|}{\cellcolor{blue!25} \textbf{0.651}} &
  \multicolumn{1}{c|}{\cellcolor{blue!25} \textbf{0.882}} &
  \multicolumn{1}{c|}{\cellcolor{blue!25} \textbf{0.508}} &
  \multicolumn{1}{c|}{\cellcolor{blue!25} \textbf{0.845}} &
  \multicolumn{1}{c|}{\cellcolor{blue!25} \textbf{0.412}}
   &
  \multicolumn{1}{c|}{\cellcolor{blue!25} \textbf{0.937}} &
  \multicolumn{1}{c|}{\cellcolor{blue!25} \textbf{0.650}} &
  \multicolumn{1}{c|}{\cellcolor{blue!25} \textbf{0.879}} &
  \multicolumn{1}{c|}{\cellcolor{blue!25} \textbf{0.506}} &
  \multicolumn{1}{c|}{\cellcolor{blue!25} \textbf{0.736}} &
  \multicolumn{1}{c|}{\cellcolor{blue!25} \textbf{0.217}}
   \\ \hline
   \textbf{\% Improvements} &
  \multicolumn{1}{c|}{\cellcolor{blue!25} \textbf{+20\%}} &
  \multicolumn{1}{c|}{\cellcolor{blue!25} \textbf{+120\%}} &
  \multicolumn{1}{c|}{\cellcolor{blue!25} \textbf{+7.7\%}} &
  \multicolumn{1}{c|}{\cellcolor{blue!25} \textbf{+19\%}} &
  \multicolumn{1}{c|}{\cellcolor{blue!25} \textbf{+18\%}} &
  \multicolumn{1}{c|}{\cellcolor{blue!25} \textbf{+132\%}}
   &
  \multicolumn{1}{c|}{\cellcolor{blue!25} \textbf{+19\%}} &
  \multicolumn{1}{c|}{\cellcolor{blue!25} \textbf{+100\%}} &
  \multicolumn{1}{c|}{\cellcolor{blue!25} \textbf{+7.8\%}} &
  \multicolumn{1}{c|}{\cellcolor{blue!25} \textbf{+21\%}} &
  \multicolumn{1}{c|}{\cellcolor{blue!25} \textbf{+13\%}} &
  \multicolumn{1}{c|}{\cellcolor{blue!25} \textbf{+83\%}}
   \\ \hline
\end{tabular}
\end{subtable}
\label{tab:defense_all}
\end{table*}

In this section, we show how much \method enhances the model stability against input perturbations across diverse datasets. We also present the effectiveness of \method under large perturbations and ablation studies of \method.

\subsection{Experimental Settings}
\label{sec:exp:settings}

\textbf{Datasets.}
We use three public recommendation datasets  that are widely used in the existing literature and span various domains.   The statistics are listed in \cref{tab:dataset}. Users with fewer than 10 interactions were filtered out.

\noindent $\bullet$ LastFM~\cite{LastFM, guo2019streaming, lei2020interactive, jagerman2019people}: This dataset consists of the music-playing history of users, represented as (user, music, timestamp). \\
\noindent $\bullet$ Foursquare~\cite{yuan2013time, ye2010location, yuan2014graph, yang2017bridging}: This dataset represents point-of-interest information, including user, location, and timestamp.\\
\noindent $\bullet$ Reddit~\cite{Reddit, JODIE, li2020dynamic, pandey2021iacn}: This dataset contains the posting history of users on subreddits, represented as (user, subreddit, timestamp).

\noindent \textbf{Target Recommender Models.}
\label{sec:exp_sequential_recommendation}
\label{sec:exp:competitors}
We aim to improve the model stability of the following state-of-the-art sequential recommenders.

\noindent $\bullet$ \textbf{\tisasrec~\cite{Tisasrec}:}
		a self-attention based model that utilizes temporal features and positional embeddings for next-item prediction.\\ 
\noindent $\bullet$ \textbf{\bert~\cite{sun2019bert4rec}:}
		a bidirectional Transformer-based model that uses masked language modeling for sequential recommendations.\\  
\noindent $\bullet$ \textbf{\lstm~\cite{hochreiter1997long}:}
		a Long Short-Term Memory (LSTM)-based model that can learn long-term dependencies using LSTM architecture. \\ 

\noindent \textbf{Evaluation Metrics.}
To quantify the stability of a recommender against perturbations, we use two Rank List Stability (RLS) metrics: RBO~\cite{RBO} and Top-$K$ Jaccard Similarity~\cite{jaccard1912distribution}. 
The metric values are between 0 to 1, and higher values are better (see \cref{sec:prelim:model_stability} for details).

To evaluate the next-item prediction accuracy, we use two popular metrics, namely, Mean Reciprocal Rank (MRR) and Recall@$K$ (typically $K=10$~\cite{JODIE,hansen2020contextual}).
The metric values are between 0 to 1, and higher values are better.
These metrics only focus on the rank of the ground-truth next item, not the ordering of all items in a rank list. Thus, they are \emph{unsuitable} to measure the model stability against input perturbations.

\noindent \textbf{Training Data Perturbation Methods.}

\noindent $\bullet$ \textbf{Random, Earliest-Random, and Latest-Random:}
Random perturbation manipulates randomly chosen interactions among all training interactions, while Earliest-Random and Latest-Random approaches perturb randomly selected interactions among the first and last $10\%$ interactions of users, respectively.

\noindent $\bullet$ \textbf{\casper~\cite{oh2022robustness}:}
\casper~\cite{oh2022robustness} is the \emph{state-of-the-art} interaction-level perturbation for sequential recommendation models. \method employs a graph-based approximation to find the most effective perturbation in the training data to alter RLS metrics. 

Although we highlight the deletion perturbation results in the paper, \method also enhances the model stability against \textbf{injection, item replacement, and mixed perturbations}. 
\method \textbf{does not know} what perturbations will be applied to the training data during its fine-tuning process.

\noindent \textbf{Baseline Fine-tuning Methods to Compare against \method.}
\label{sec:exp:baselines}

\noindent $\bullet$ \textbf{Original:} It trains a recommender model on original training data (without fine-tuning) with standard next-item prediction loss.

\noindent $\bullet$ \textbf{Random, Earliest-Random, and Latest-Random:}
The Random method perturbs 1\% of random training interactions for every epoch (either deletion, insertion, or replacement) and fine-tunes a recommender model with the perturbed data. The Earliest-Random and Latest-Random randomly perturb 1\% of interactions in the first and last 10\% (based on timestamps) of the training data and fine-tune the model with the perturbed data, respectively.

\noindent $\bullet$ \textbf{Adversarial Poisoning Training (APT)~\cite{wu2021fight}:} 
APT is the \textbf{state-of-the-art} adversarial training method that fine-tunes a recommendation model using perturbed training data, including fake user profiles. Since it only works for matrix factorization-based models, we replace the fake user generation part with the Data-free~\cite{yue2021black} model and fine-tune the recommender with the perturbed data.
	
\noindent $\bullet$ \textbf{ACAE~\cite{yuan2019adversarial}:} 
ACAE is another \textbf{state-of-the-art} adversarial training model that adds gradient-based
noise (found by the fast gradient method~\cite{FGSM}) to the model parameters while fine-tuning a recommendation model. To adapt it to our setting, we add noise to the input sequence embeddings instead of model parameters.

Note that \casper~\cite{oh2022robustness} is an input perturbation method and cannot be compared with \method directly.
We also exclude several baselines if they are designed for multimodal recommender systems~\cite{anelli2021study, tang2019adversarial} which require additional input data like images, or if they do not provide fine-tuning mechanisms~\cite{yue2022defending, tan2023towards, wu2021triple, MSAP, du2018enhancing, zhang2021data}.

\noindent \textbf{Experimental Setup.}
\method is implemented in Python and PyTorch library and tested in the NVIDIA DGX machine that has 5 NVIDIA A100 GPUs with 80GB memory.
We use the first 90\% of interactions of each user for training and validation, and the rest of the interactions are used for testing.
For \method, we use the following hyperparameters found by grid searches on validation data. 
Please refer to the Appendix for detailed hyperparameter values we tested.
The sampling ratio of interactions for perturbation simulations is set to 1\%, and we sample the top-200 items for regularization, and the regularization coefficients $\lambda, \lambda_1, \lambda_2$ are set to $1.0$, $0.1$, and $0.1$, respectively. We assign 50 epochs for fine-tuning.
The maximum training epoch is set to 100, a learning rate is set to 0.001, and the embedding dimension is set to 128. 
For all recommendation models, the maximum sequence length per user is set to 50. 
We also perturb 0.1\% of training interactions.
We repeat all experiments three times with different random seeds and report average values of RLS and next-item metrics.
To measure statistical significance, we use the one-tailed t-test.

\begin{table}[t!]
\caption{\textit{Next-item prediction performance of \method on LastFM and Foursquare datasets (no perturbations).} \method successfully preserves or enhances next-item metrics of all recommendation models. Results with * indicate statistical significance (p-value < 0.05).  
}
\begin{subtable}[h]{1.0\linewidth}
\centering
\caption{LastFM Dataset}
\begin{tabular}{|c|cccccc|}
\hline
\textbf{\begin{tabular}[c]{@{}c@{}}Recommenders \end{tabular}} &
  \multicolumn{2}{c|}{\textbf{\tisasrec~\cite{Tisasrec}}} &
  \multicolumn{2}{c|}{\textbf{\bert~\cite{sun2019bert4rec}}} &
  \multicolumn{2}{c|}{\textbf{\lstm~\cite{hochreiter1997long}}} \\ \hline
\textbf{\begin{tabular}[c]{@{}c@{}}Next-item Metrics \end{tabular}} &
  \multicolumn{1}{c|}{\textbf{MRR}} &
  \multicolumn{1}{c|}{\textbf{\begin{tabular}[c]{@{}c@{}}Recall \\ @10 \end{tabular}}} &
  \multicolumn{1}{c|}{\textbf{MRR}} &
  \multicolumn{1}{c|}{\textbf{\begin{tabular}[c]{@{}c@{}}Recall \\ @10 \end{tabular}}} &
  \multicolumn{1}{c|}{\textbf{MRR}} &
  \multicolumn{1}{c|}{\textbf{\begin{tabular}[c]{@{}c@{}}Recall \\ @10 \end{tabular}}} \\ \hline
\textbf{\begin{tabular}[c]{@{}c@{}}Original Model \end{tabular}} &
  \multicolumn{1}{c|}{\cellcolor{red!25} 0.154} &
  \multicolumn{1}{c|}{\cellcolor{red!25} 0.277} &
  \multicolumn{1}{c|}{\cellcolor{red!25} 0.158} &
  \multicolumn{1}{c|}{\cellcolor{red!25} 0.318} &
  \multicolumn{1}{c|}{\cellcolor{red!25} 0.156} &
  \multicolumn{1}{c|}{\cellcolor{red!25} 0.235} 
   \\ \hline
\textbf{\begin{tabular}[c]{@{}c@{}}\method (Proposed) \end{tabular}} &
  \multicolumn{1}{c|}{\cellcolor{blue!25} \textbf{0.164*}} &
  \multicolumn{1}{c|}{\cellcolor{blue!25} \textbf{0.299*}} &
  \multicolumn{1}{c|}{\cellcolor{blue!25} \textbf{0.168*}} &
  \multicolumn{1}{c|}{\cellcolor{blue!25} \textbf{0.343*}} &
  \multicolumn{1}{c|}{\cellcolor{blue!25} \textbf{0.169*}} &
  \multicolumn{1}{c|}{\cellcolor{blue!25} \textbf{0.252*}}
   \\ \hline
\end{tabular}
\end{subtable} 
\\
\begin{subtable}[h]{1.0\linewidth}
\centering
\caption{Foursquare Dataset}
\begin{tabular}{|c|cccccc|}
\hline
\textbf{\begin{tabular}[c]{@{}c@{}}Recommenders \end{tabular}} &
  \multicolumn{2}{c|}{\textbf{\tisasrec~\cite{Tisasrec}}} &
  \multicolumn{2}{c|}{\textbf{\bert~\cite{sun2019bert4rec}}} &
  \multicolumn{2}{c|}{\textbf{\lstm~\cite{hochreiter1997long}}} \\ \hline
\textbf{\begin{tabular}[c]{@{}c@{}}Next-item Metrics \end{tabular}} &
  \multicolumn{1}{c|}{\textbf{MRR}} &
  \multicolumn{1}{c|}{\textbf{\begin{tabular}[c]{@{}c@{}}Recall \\ @10 \end{tabular}}} &
  \multicolumn{1}{c|}{\textbf{MRR}} &
  \multicolumn{1}{c|}{\textbf{\begin{tabular}[c]{@{}c@{}}Recall \\ @10 \end{tabular}}} &
  \multicolumn{1}{c|}{\textbf{MRR}} &
  \multicolumn{1}{c|}{\textbf{\begin{tabular}[c]{@{}c@{}}Recall \\ @10 \end{tabular}}} \\ \hline
\textbf{\begin{tabular}[c]{@{}c@{}}Original Model \end{tabular}} &
  \multicolumn{1}{c|}{\cellcolor{red!25} 0.082} &
  \multicolumn{1}{c|}{\cellcolor{red!25} 0.137} &
  \multicolumn{1}{c|}{\cellcolor{red!25} 0.162} &
  \multicolumn{1}{c|}{\cellcolor{red!25} 0.267} &
  \multicolumn{1}{c|}{\cellcolor{red!25} 0.096} &
  \multicolumn{1}{c|}{\cellcolor{red!25} 0.166} 
   \\ \hline
\textbf{\begin{tabular}[c]{@{}c@{}}\method (Proposed) \end{tabular}} &
  \multicolumn{1}{c|}{\cellcolor{blue!25} \textbf{0.088}} &
  \multicolumn{1}{c|}{\cellcolor{blue!25} \textbf{0.143}} &
  \multicolumn{1}{c|}{\cellcolor{blue!25} \textbf{0.175*}} &
  \multicolumn{1}{c|}{\cellcolor{blue!25} \textbf{0.283*}} &
  \multicolumn{1}{c|}{\cellcolor{blue!25} \textbf{0.102}} &
  \multicolumn{1}{c|}{\cellcolor{blue!25} \textbf{0.174}}
   \\ \hline
\end{tabular}
\end{subtable}
\label{tab:defense_next_item_metrics}
\end{table}

\begin{figure}[t!]
    \begin{subfigure}{0.48\linewidth}
    \centering
    \includegraphics[width=6.5cm]{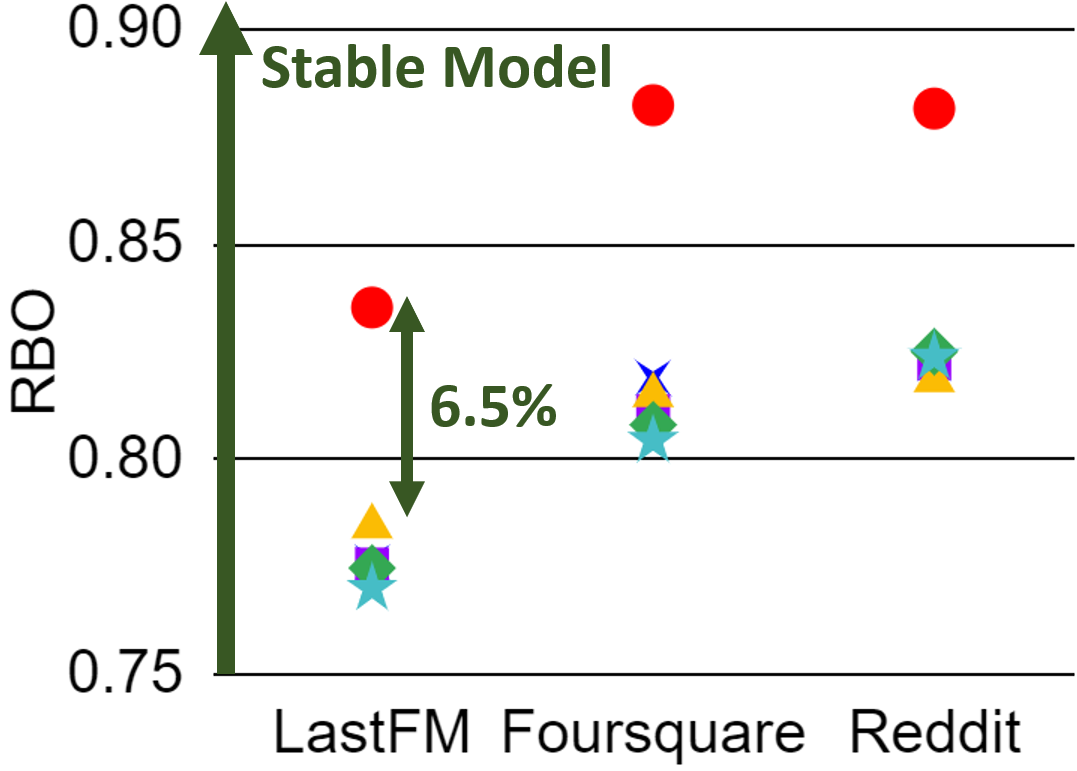}
    \captionsetup{justification=centering}
    \caption{RBO on Random Perturbation}
    \label{fig:all_random_rbo}
    \end{subfigure}
    \begin{subfigure}{0.48\linewidth}
    \centering
    \includegraphics[width=6.5cm]{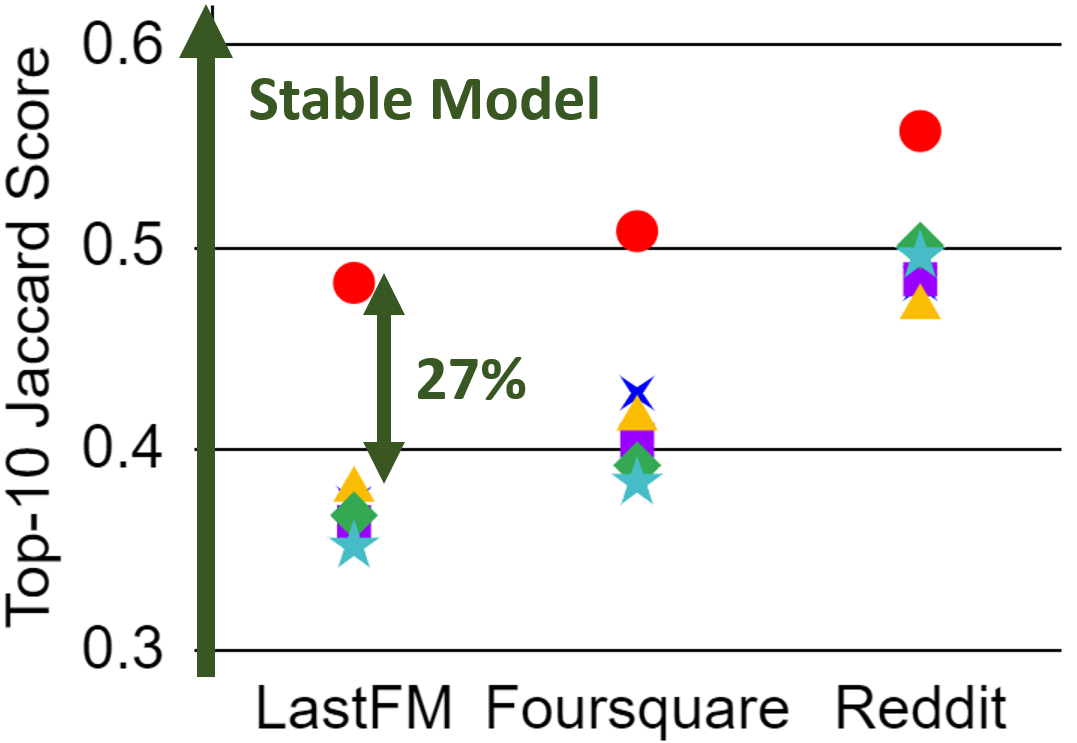}
    \captionsetup{justification=centering}
    \caption{Jaccard on Random Perturbation}
    \label{fig:all_random_jaccard}
    \end{subfigure}
    \\
    \begin{subfigure}{0.48\linewidth}
    \centering
    \includegraphics[width=6.5cm]{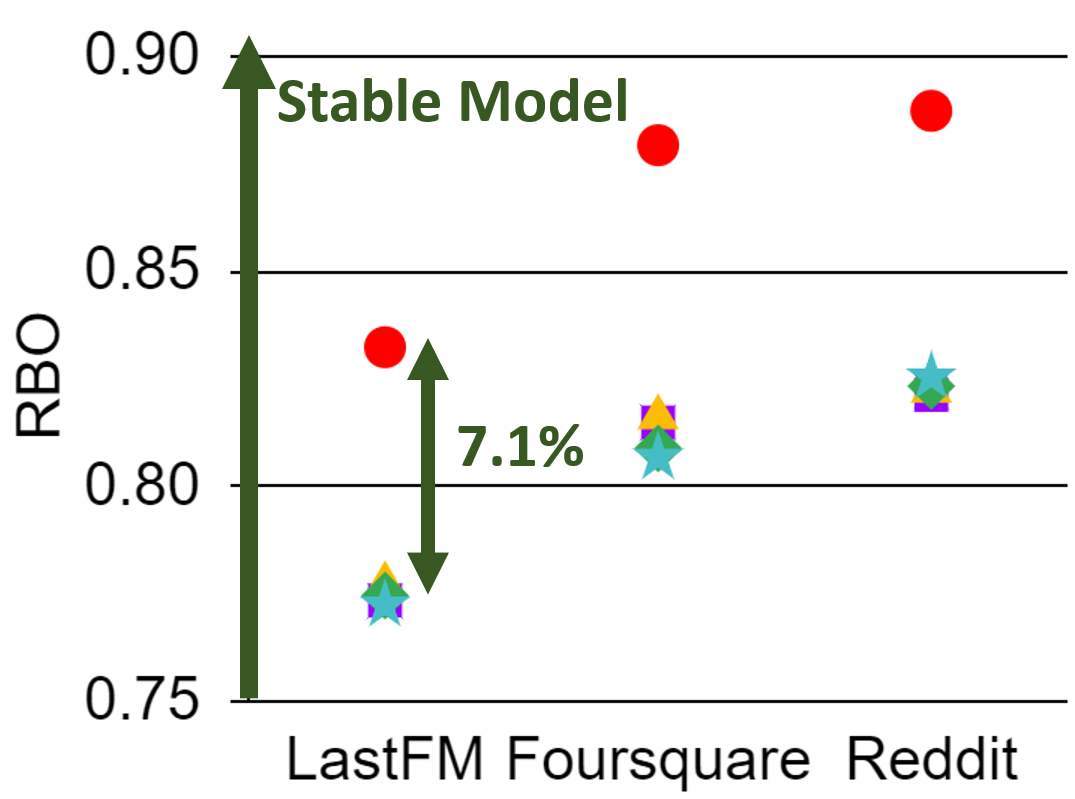}
    \captionsetup{justification=centering}
    \caption{RBO on \casper Perturbation}
    \label{fig:all_casper_rbo}
    \end{subfigure}
    \begin{subfigure}{0.48\linewidth}
    \centering
    \includegraphics[width=6.5cm]{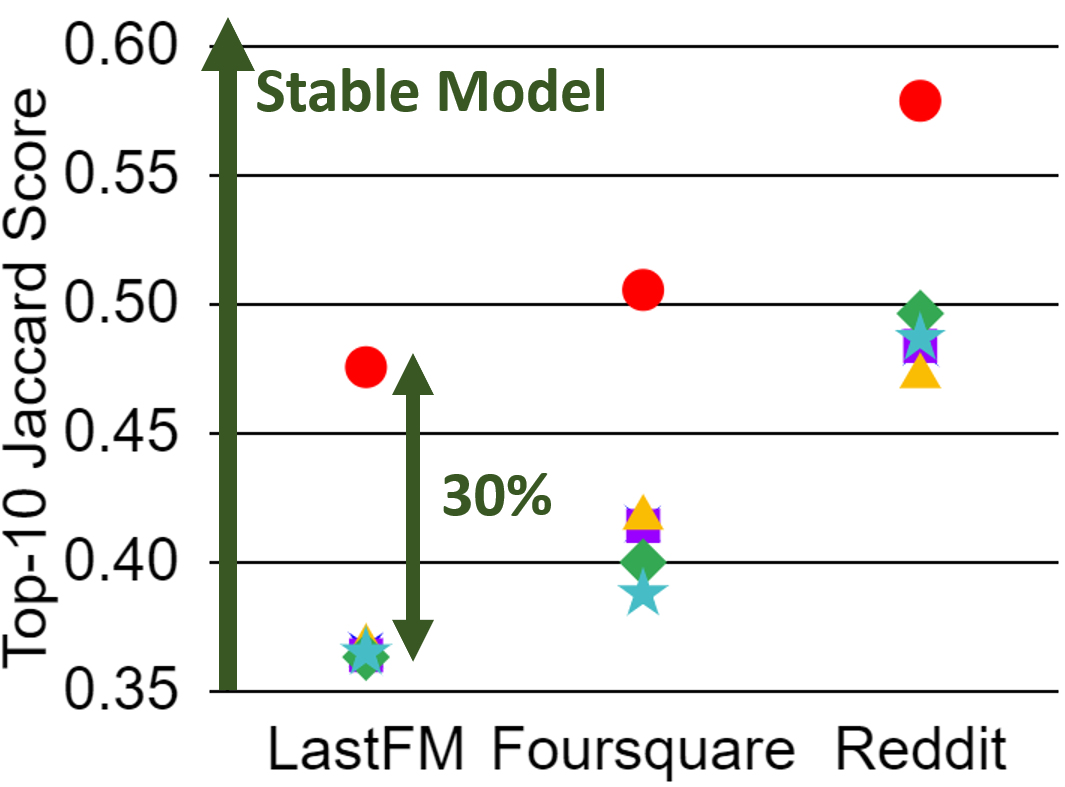}
    \captionsetup{justification=centering}
    \caption{Jaccard on \casper Perturbation}
    \label{fig:all_casper_jaccard}
    \end{subfigure}
    \begin{subfigure}{1.0\linewidth}
    \centering
    \includegraphics[width=1.0\linewidth]{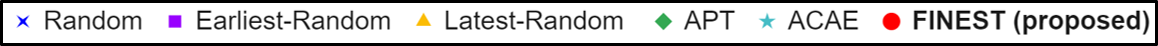}
    \end{subfigure}
    \caption{\textit{
    Stability of the \bert model fine-tuned with diverse methods against random and \casper~\cite{oh2022robustness} deletion perturbations across different datasets.} 
    \method generates the most stable model against both perturbations as per RBO and Top-10 Jaccard Similarity.
    }
	\label{fig:all_dataset}
\end{figure}

\subsection{Effectiveness of \method}
\label{sec:exp:best_defense}

\noindent  \textbf{Fine-tuning method comparison on the LastFM and Foursquare datasets.}
In \cref{tab:defense_all}, we compare the performance of all fine-tuning methods on all three recommendation models against random and \casper~\cite{oh2022robustness} deletion perturbations on the LastFM and Foursquare datasets. We highlight the results on these two datasets as they are the top-2 largest ones as per the number of interactions. 
Each column shows the \textbf{{\color{red!50}original}} method without any fine-tuning and the \textbf{{\color{blue!50}best}} fine-tuning method with the highest RLS value. 

Our proposed fine-tuning method \method \textbf{outperforms} all of the baselines across all recommender systems, with statistical significance in all cases (p-values < 0.05). 
\method demonstrates significant improvements in RLS metrics compared to the results of original training and baselines. For instance, on the \lstm model (the most susceptible one against \casper perturbation), \method shows at least 12.4\% RBO and 82.5\% Jaccard improvements versus the best baseline. Even on the \bert model (the most stable one against \casper perturbation), \method still exhibits at least 7.5\% RBO and 21.2\% Jaccard score boosts versus the best baseline. 
Baseline fine-tuning methods have limitations in improving model stability since they do not incorporate the rank list preservation component in their fine-tuning.
We observe the same trend for other types of perturbations and datasets (e.g., see \cref{fig:all_dataset}).
Thus, with \method, the state-of-the-art recommender systems can generate stable rank lists even after perturbations.

\noindent  \textbf{Impact of \method on next-item prediction accuracy.}
Fine-tuning the recommender with randomly sampled perturbations can increase or preserve the next-item prediction accuracy (e.g., MRR and Recall@10). \textit{This is due to the implicit data augmentation and cleaning effect from the perturbed training examples}. 
We validate the improvements of both next-item prediction performance and model stability  in \cref{tab:defense_next_item_metrics} (on the LastFM and Foursquare datasets, with no perturbations).
\cref{tab:defense_next_item_metrics}  demonstrates that \method can boost the model stability without sacrificing its next-item prediction accuracy with statistical significance in most cases. \tisasrec shows relatively lower next-item prediction performance as it is optimized for sampled metrics, which computes ranking with negative items during the test. For more details, please refer to the paper~\cite{krichene2022sampled}. Meanwhile, the other models are optimized over all items.

\noindent \textbf{Fine-tuning method comparison on different datasets.}
We further evaluate the effectiveness of \method versus the baselines on the \bert model (which shows high accuracy and stability) across various datasets. 
The results are shown in \cref{fig:all_dataset}.  
We find that \method exhibits the highest model stability with statistical significance (p-values < 0.05) across all datasets and two perturbations, as per both RLS metrics. 
For instance, on the LastFM dataset (the largest one in terms of the number of interactions), \method offers at least $6.5$\% stability improvements compared to baselines in terms of RBO and at least $27$\% in top-10 Jaccard similarity.

\begin{figure}[t!]
    \centering
    \includegraphics[width=9cm]{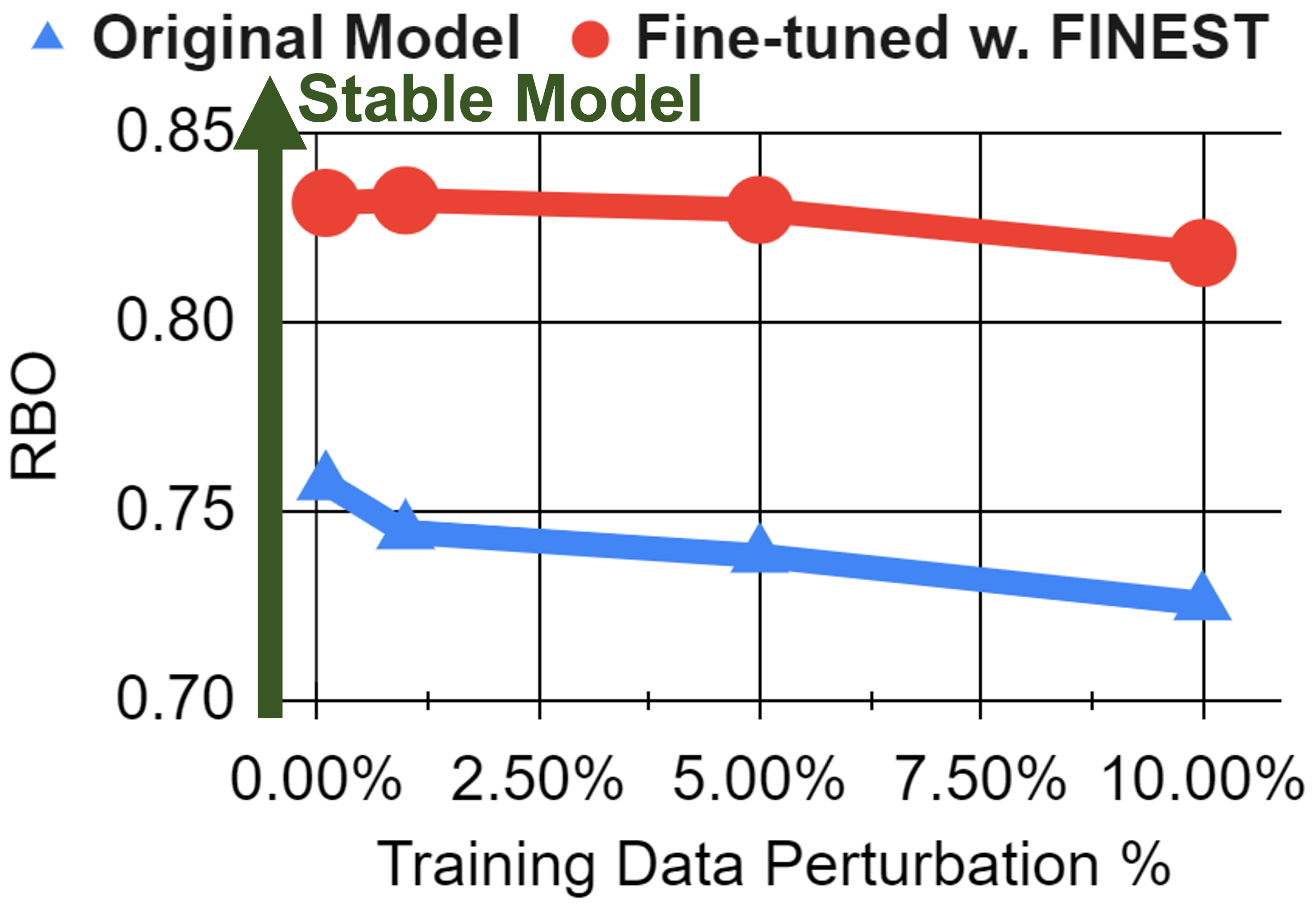}
    \caption{\textit{Stability of \bert fine-tuned with and without \method as per the number of input perturbations on the LastFM dataset.} 
    }
   \label{fig:defense_scalability}
	\label{fig:runtime_and_attack_scalability}
\end{figure}

\subsection{Model Stability against Large Perturbations}
We test how much the RLS metric of \method changes with respect to the number of input perturbations, since more perturbations will naturally lower the model stability further. \cref{fig:defense_scalability} shows the RBO scores of the \bert model trained with and without \method on the LastFM dataset against \casper deletion perturbations while varying the perturbation scale from 0.1\% to 10\%. \method provides significant improvements in the model stability compared to the original model across all perturbation scales.  

\subsection{Scalability of \method}
The time and space complexities of \method scale near-linearly to the number of interactions and items in a dataset. 
Empirically, we also confirm that \method can enhance the stability of recommenders on \textbf{large-scale datasets} such as MovieLens-10M~\cite{movielens} (72K users, 10K items, and 10M interactions; runtime = 6.6 hours)  or Steam~\cite{kang2018self} (2.6M users, 15K items, and 7.8M interactions; runtime = 9.7 hours) dataset.
For instance,  \bert with \method shows a 12\% improvement in the RBO metric ($0.763 \rightarrow 0.853$) compared to the original \bert on the Steam dataset with random perturbations.

\begin{figure}[t!]
    \includegraphics[width=10cm]{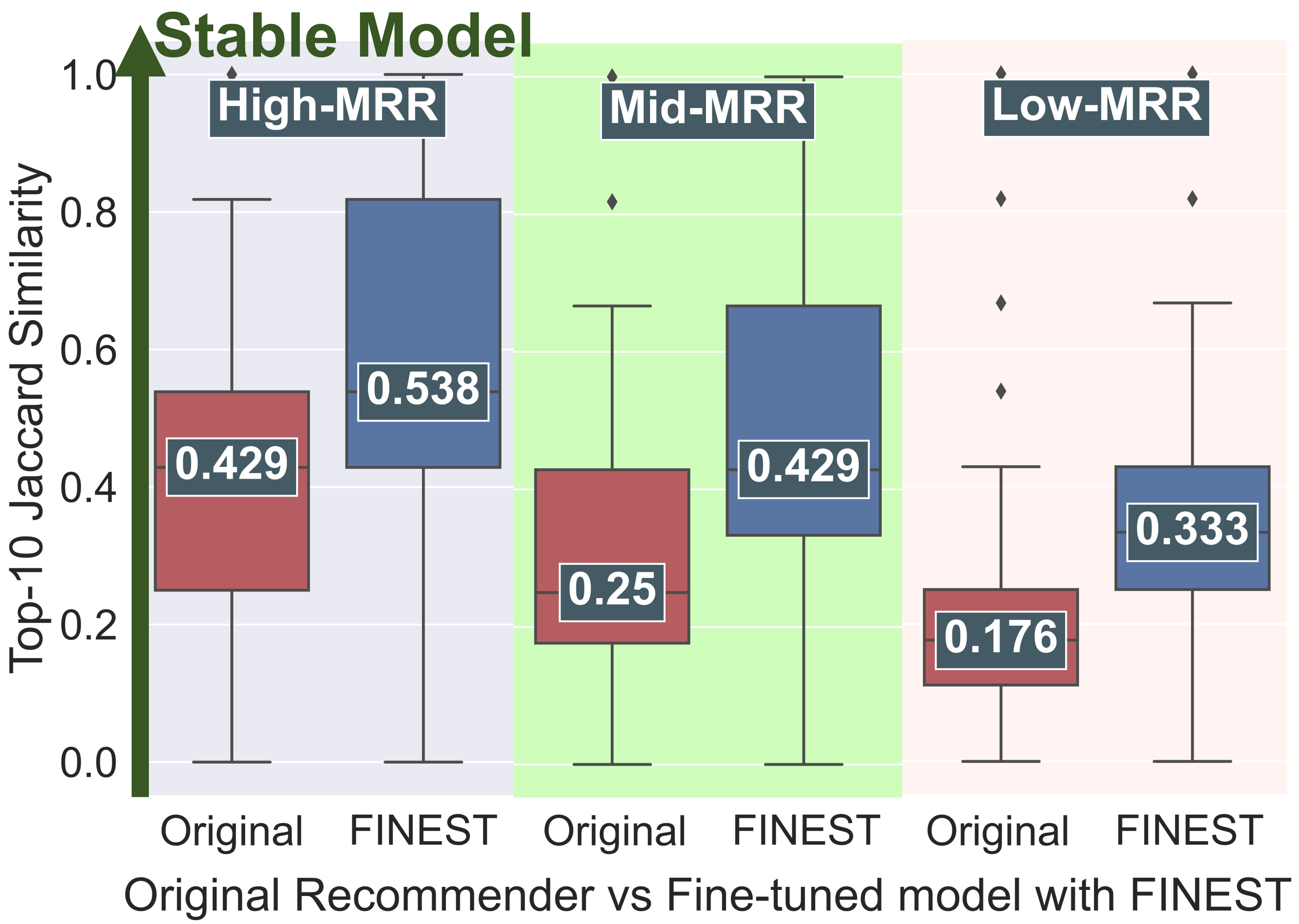}
    \caption{
    \method enhances the stability of recommendations across all user groups with different next-item prediction accuracies. These results are for recommendations from the \bert model on the LastFM dataset with respect to the \casper~\cite{oh2022robustness} perturbation. 
    }
	\label{fig:fairness}
\end{figure}

\subsection{Effectiveness on Different User Groups}
It has been shown that input perturbations can disproportionately affect users' recommendation results~\cite{oh2022robustness}.
A stability analysis result of the \bert model on the LastFM dataset against the \casper perturbation supports that observation. As shown in \cref{fig:fairness}, a \emph{low-accuracy} user group (with the lowest 20\% MRR metric on average among all users) receives unstable recommendations compared to the \emph{high-accuracy} group. This can raise fairness concerns between user groups similar to ``the rich get richer'' problem. \method can mitigate this issue by enhancing the stability of the model, thereby narrowing the relative stability difference between the two groups (e.g., $143\%$ without \method $\rightarrow$ 62\% with \method).

\begin{table}[t!]
\caption{Ablation study of the key components of \method.}
\begin{tabular}{|c|cc|cc|}
\hline
 &
  \multicolumn{2}{c|}{\textbf{RLS metrics}} &
  \multicolumn{2}{c|}{\textbf{Next-item metrics}} \\ \hline
\textbf{Fine-tuning Methods / Metrics} &
  \multicolumn{1}{c|}{\textbf{RBO}} &
  \textbf{Jaccard} &
  \multicolumn{1}{c|}{\textbf{MRR}} &
  \textbf{Recall} \\ \hline
\textbf{\begin{tabular}[c]{@{}c@{}}Original (no fine-tune) \end{tabular}} &
  \multicolumn{1}{c|}{\cellcolor{red!25} 0.7538} & \cellcolor{red!25} 
  0.3157 &
  \multicolumn{1}{c|}{\cellcolor{red!25} 0.1580} & \cellcolor{red!25} 
  0.3181 \\ \hline
  \textbf{\begin{tabular}[c]{@{}c@{}}\method without \\ Perturbation Simulation\end{tabular}} &
  \multicolumn{1}{c|}{0.8124} &
  0.4349 &
   \multicolumn{1}{c|}{0.1634} &
  0.3324 \\ \hline
  \textbf{\begin{tabular}[c]{@{}c@{}}\method without \\ Top-$K$ Regularization\end{tabular}} &
  \multicolumn{1}{c|}{0.7734} &
  0.3662 &
  \multicolumn{1}{c|}{0.1665} &
  0.3366 \\ \hline
  \textbf{\begin{tabular}[c]{@{}c@{}}\method without \\ First Regularization Term \end{tabular}} &
  \multicolumn{1}{c|}{0.8268} &
   0.4658 &
   \multicolumn{1}{c|}{0.1663} & 0.3379
   \\ \hline
  \textbf{\begin{tabular}[c]{@{}c@{}}\method without \\ Second Regularization Term\end{tabular}} &
  \multicolumn{1}{c|}{0.7955} &
  0.4245 &
   \multicolumn{1}{c|}{\cellcolor{blue!25} \textbf{0.1713}} &
  \cellcolor{blue!25} \textbf{0.3513} \\ \hline
\textbf{\method (proposed)} &
  \multicolumn{1}{c|}{\cellcolor{blue!25} \textbf{0.8324}} &
  \cellcolor{blue!25} \textbf{0.4757} &
   \multicolumn{1}{c|}{0.1682} &
  0.3434 \\ \hline
\end{tabular}
\label{tab:ablation_study}
\end{table}

\subsection{Ablation Studies of \method}
\label{sec:exp:ablation_study}
We verify the contributions of the perturbation simulation and rank-preserving regularization of \method by measuring the model stability after removing each component. 
\cref{tab:ablation_study} shows the ablation study results on the \bert model and LastFM dataset against \casper deletion perturbations in terms of RLS and next-item metrics. 
We observe that all variants of \method outperform the original training (without fine-tuning) in all metrics. Among the variants, we see that the model without the perturbation simulation performs better than the model without the regularization, implying that the top-$K$ regularization has a higher impact on enhancing the model stability. 
Regarding the regularization function, the ``score-preserving'' component (second term in \cref{eq:topK_regularization_single}) is more effective in terms of RLS metrics than the ``ordering-preserving'' component (first term in \cref{eq:topK_regularization_single}).
In summary, having both components of \method together results in the highest model stability. 

\subsection{Hyperparameter Sensitivity of \method}
\begin{figure}[t!]
	\centering
	\begin{subfigure}[t]{0.24\linewidth}
		\includegraphics[width=3.7cm]{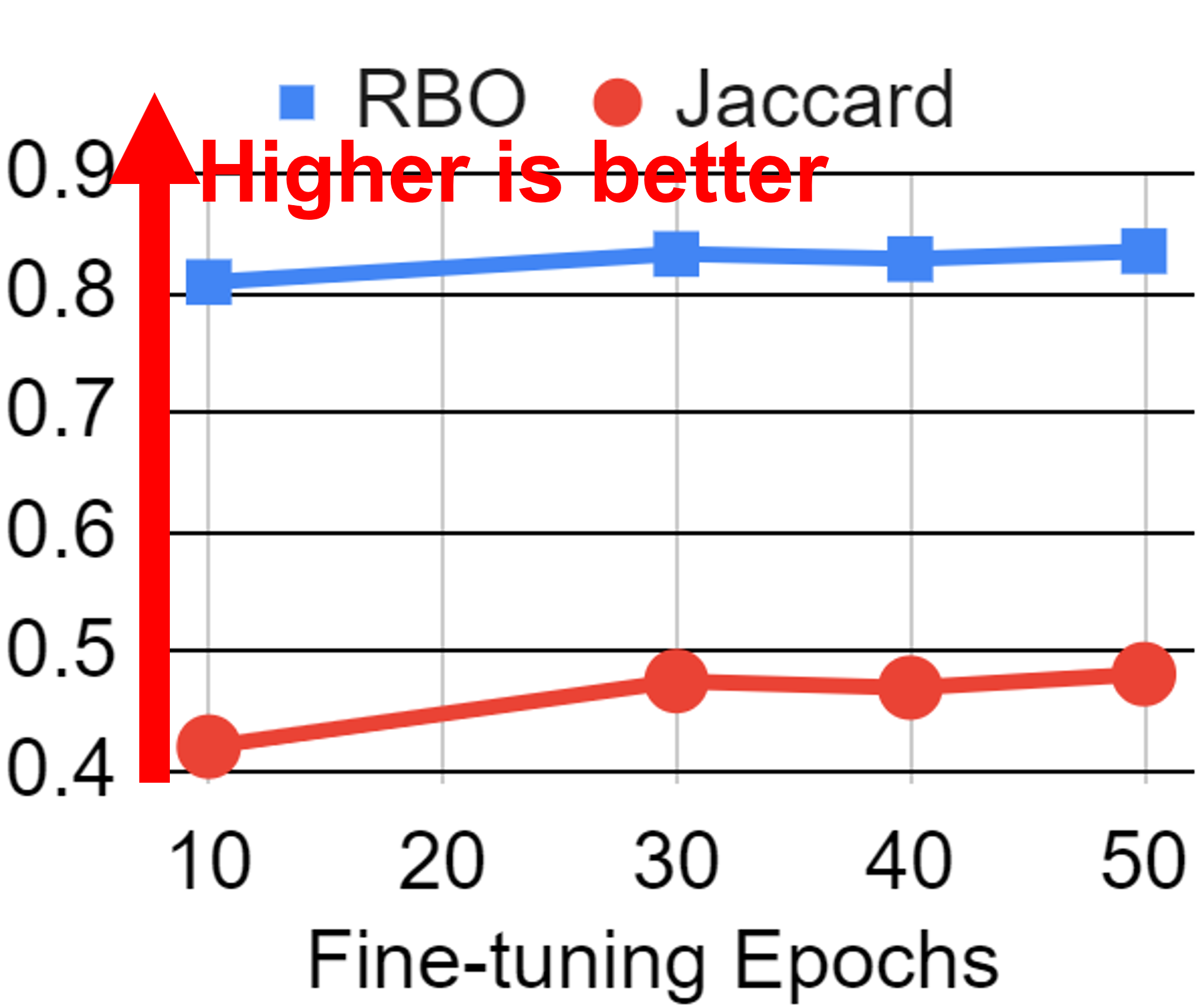}
	\end{subfigure}
	\begin{subfigure}[t]{0.24\linewidth}
		\includegraphics[width=3.7cm]{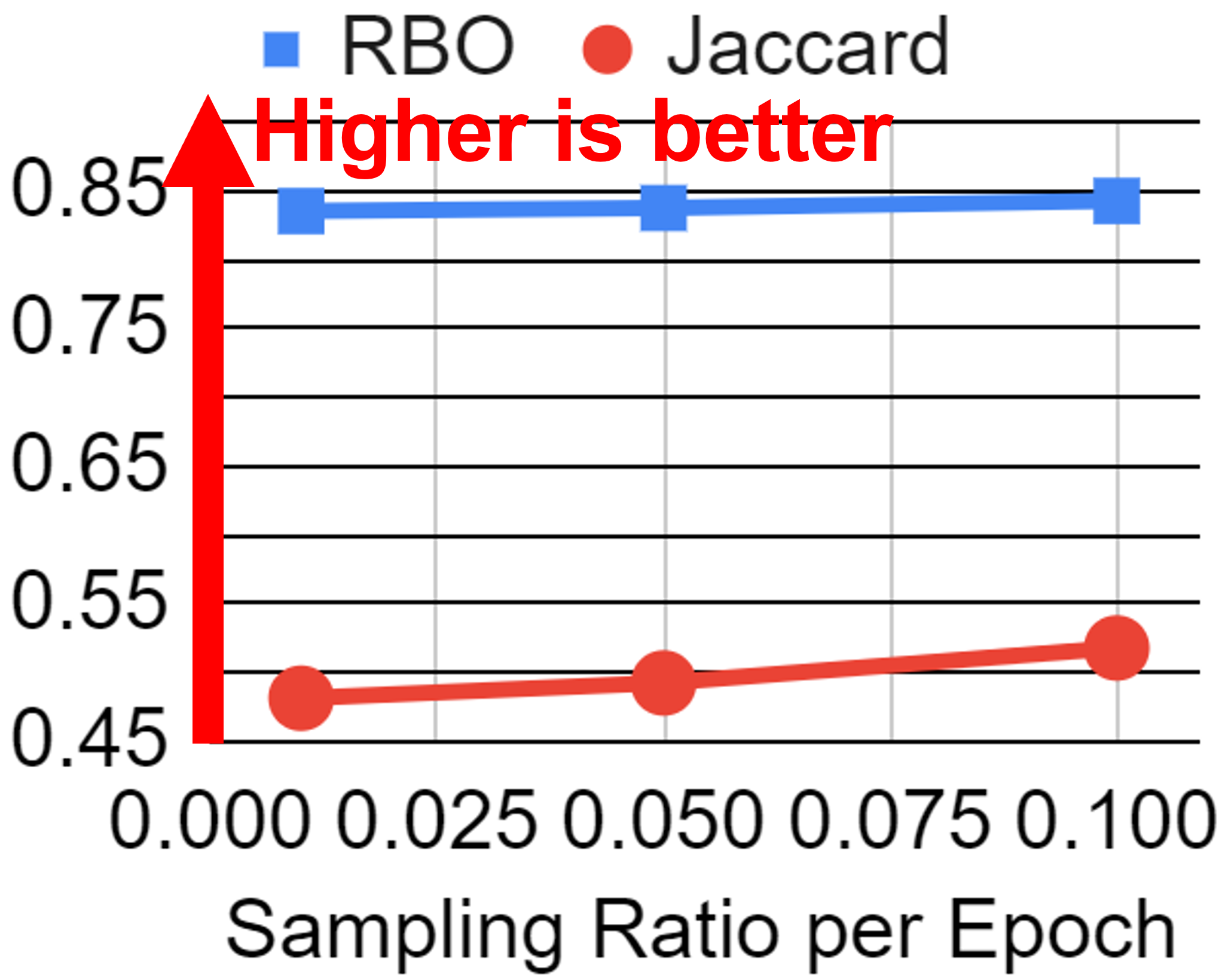}
	\end{subfigure}
	\begin{subfigure}[t]{0.23\linewidth}
		\includegraphics[width=3.7cm]{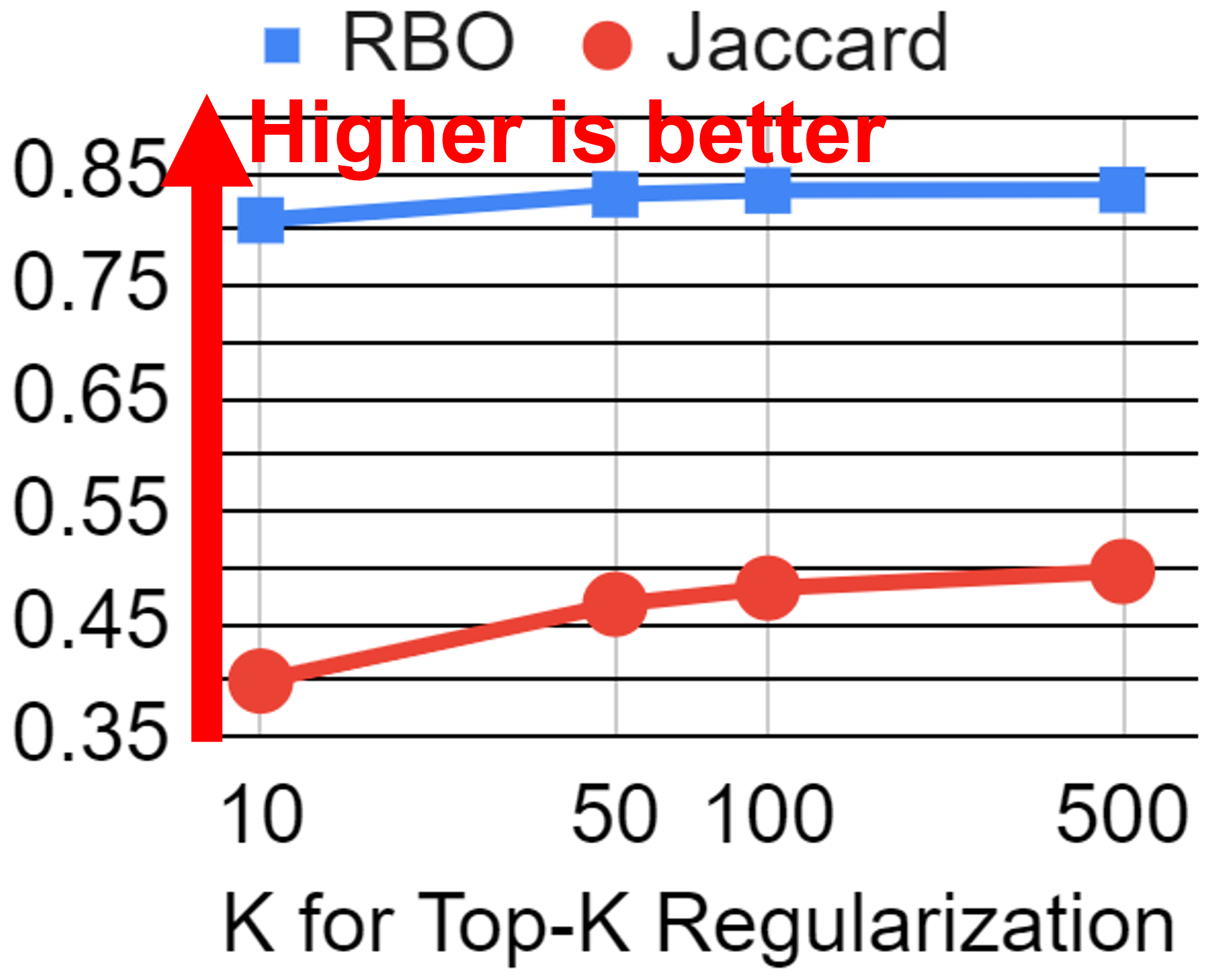}
	\end{subfigure}
	\begin{subfigure}[t]{0.23\linewidth}
		\includegraphics[width=3.7cm]{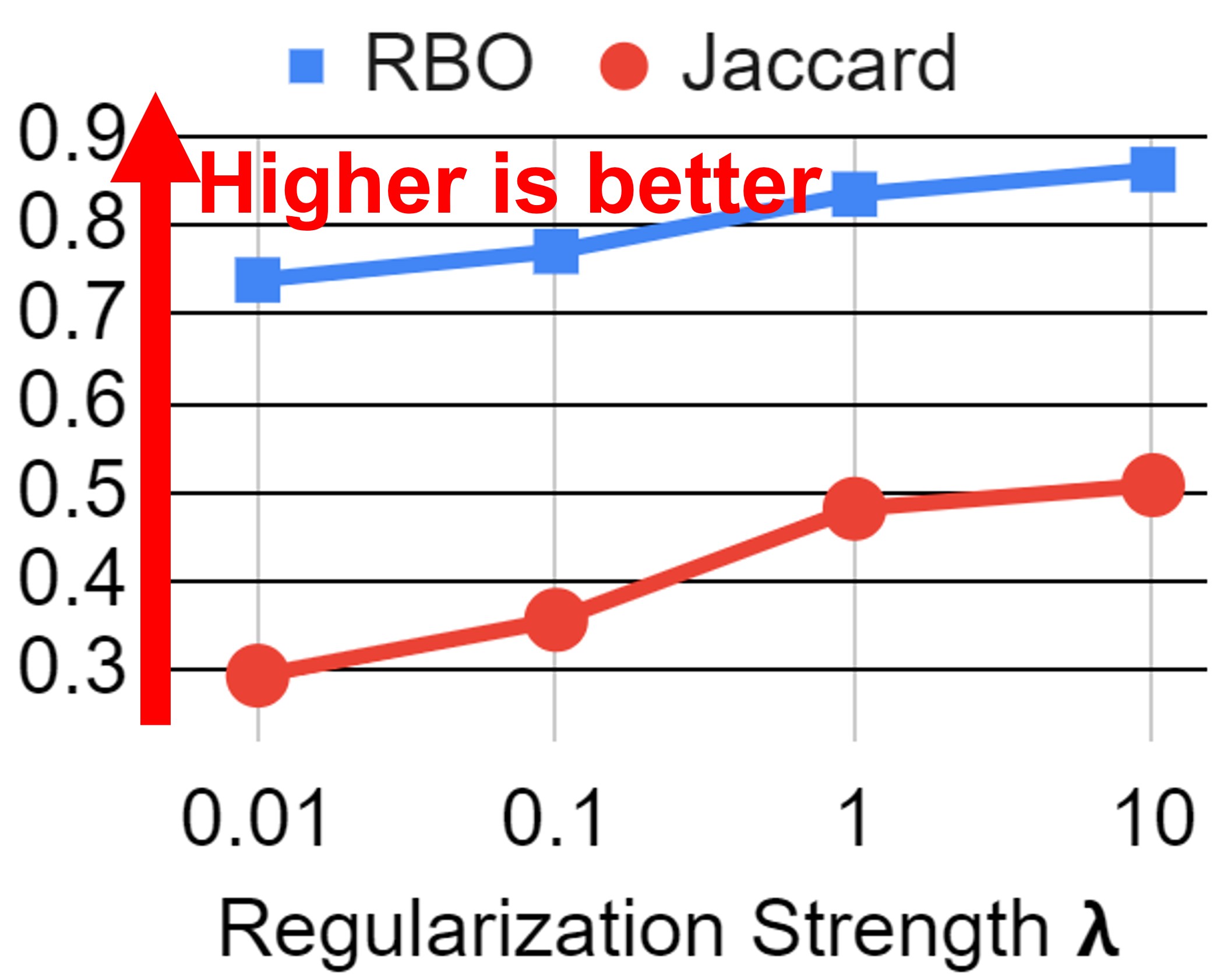}
	\end{subfigure}
	\\
 	\begin{subfigure}[t]{0.24\linewidth}
		\includegraphics[width=3.7cm]{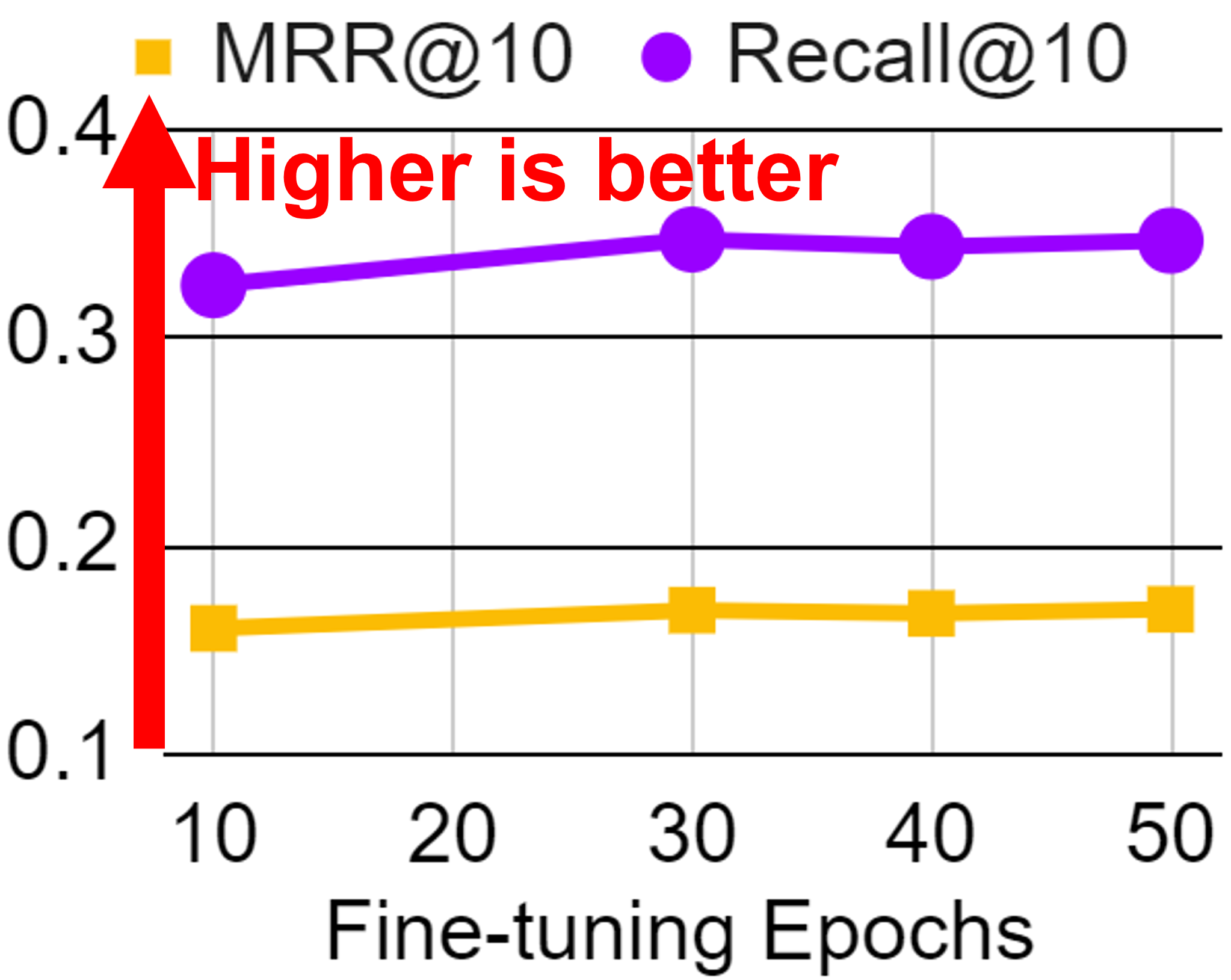}
	\end{subfigure}
	\begin{subfigure}[t]{0.24\linewidth}
		\includegraphics[width=3.7cm]{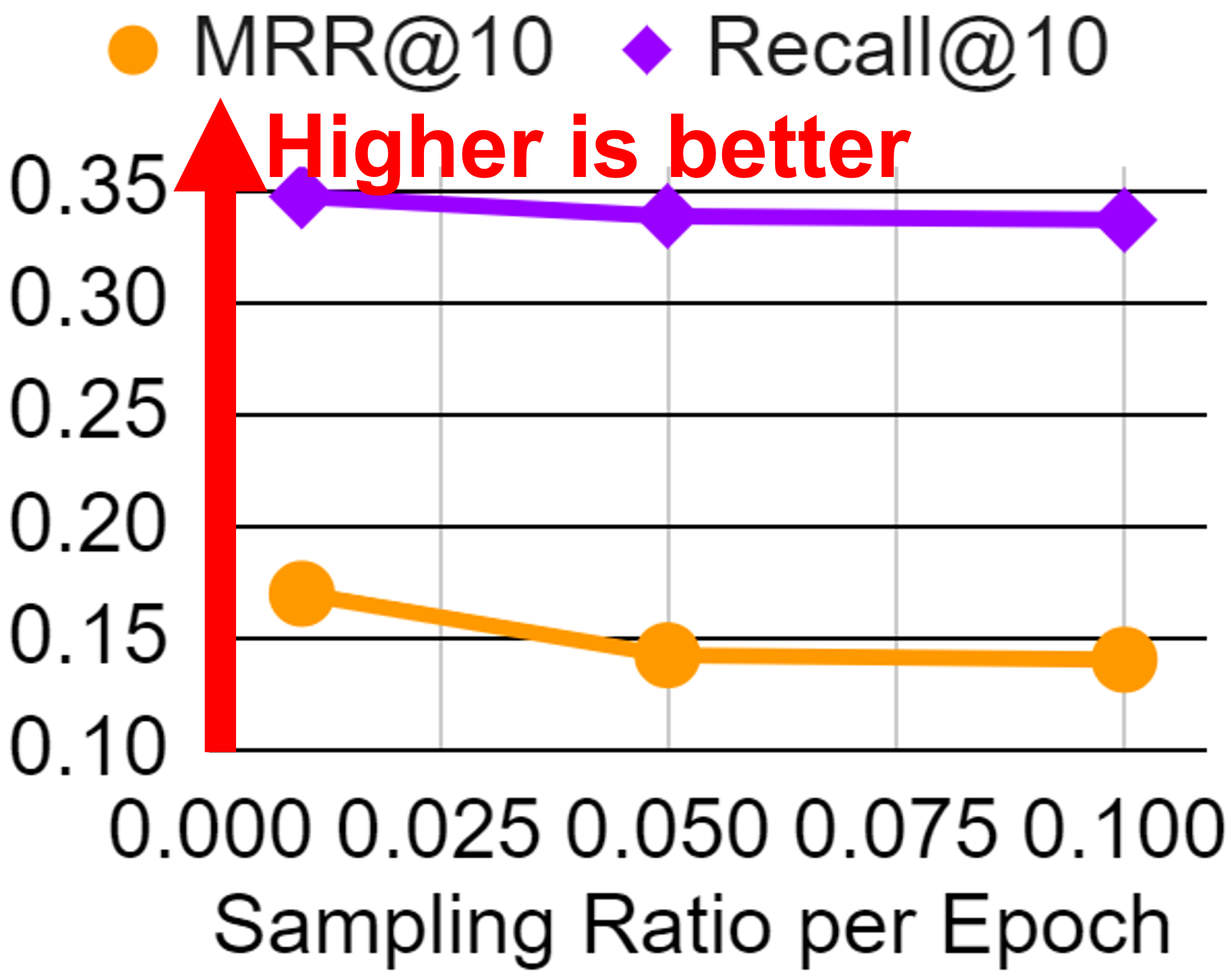}
	\end{subfigure}
	\begin{subfigure}[t]{0.23\linewidth}
		\includegraphics[width=3.7cm]{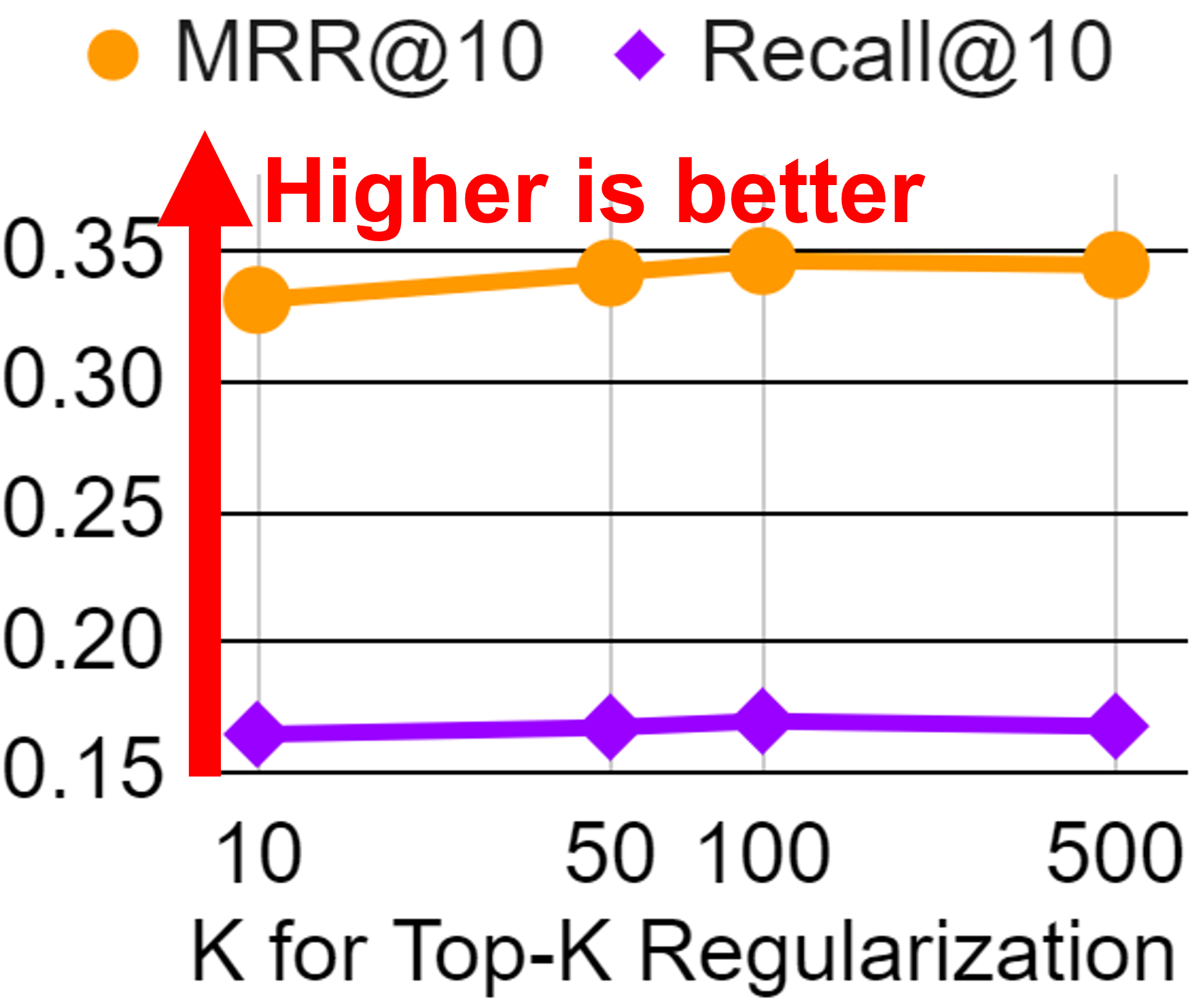}
	\end{subfigure}
	\begin{subfigure}[t]{0.23\linewidth}
		\includegraphics[width=3.7cm]{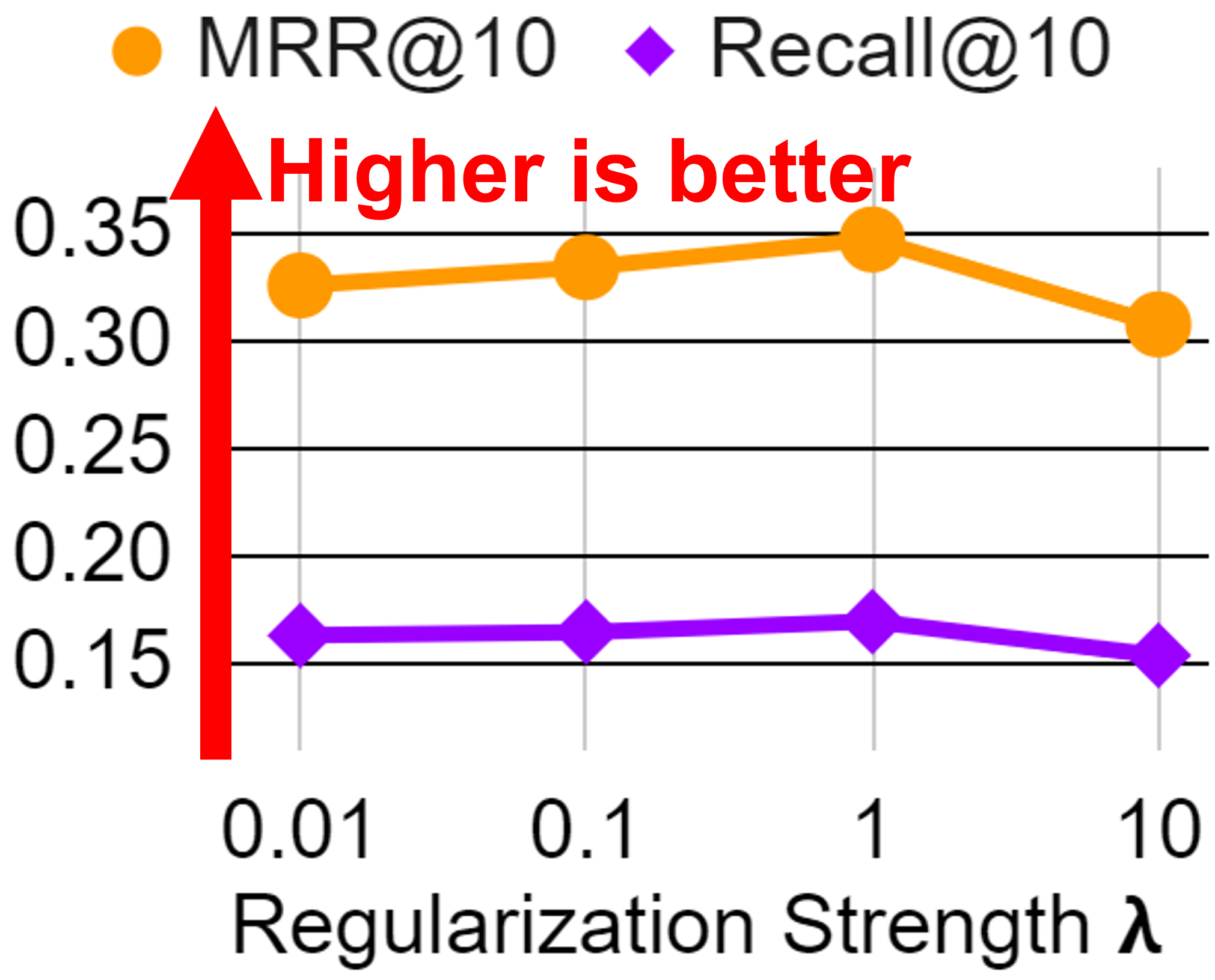}
	\end{subfigure}
	\caption{\textit{Hyperparameter sensitivity of \method on the \bert model and LastFM dataset against \casper deletion perturbations.} 
	}
	\label{fig:hyperparameter}
\end{figure}

Figure~\ref{fig:hyperparameter} exhibits the hyperparameter sensitivity of \method with respect to RLS and next-item metrics on the \bert model and LastFM dataset against \casper deletion perturbations. 
We change one hyperparameter while fixing all the others to the default values stated in Section~\ref{sec:exp:settings}.
We find that both metrics improve as fine-tuning continues, and the improvements saturate after sufficient epochs (e.g., 50) of fine-tuning are done. 
Regarding the sampling ratio, we observe the trade-off between RLS and next-item metrics as the ratio increases. In practice, a small value (e.g., 1\%) is preferred as a high value can hurt the next-item metrics. 
A medium number of top-K items (e.g., 100) is best for \method since a small value can have a minor impact on preserving the rank lists, and a large value can reduce the scalability of \method. 
Finally, a medium value of $\lambda$ (e.g., 1) leads to high RLS and next-item metrics, as a small value limits the effect of the ranking-preserving regularization, while a large value can lead to inaccurate next-item predictions. 

	\section{Discussion \& Conclusion}
	\label{sec:discussion}
	\noindent \textbf{Why is Fine-tuning Selected over Retraining?}
One may wonder whether training with \method from scratch (instead of fine-tuning) is sufficient for achieving high model stability. There are two key reasons why fine-tuning is preferred. 
First, existing literature in recommender systems~\cite{yuan2019adversarial, wu2021fight} has demonstrated that fine-tuning mechanisms should be applied when the given model starts to overfit~\cite{he2018adversarial} for the best model robustness, not when it is still in the early stage. Thus, it is better to apply the fine-tuning method \method after the model has been trained sufficiently, not from the beginning.
Second, our fine-tuning process requires the rank lists of all training instances as a reference for the regularization. 
If the model is not fully trained as per next-item prediction accuracy, the reference lists will not be optimal. A pre-trained recommendation model ensures that appropriate reference lists are used. 
Third, fine-tuning techniques can be applied to existing pre-trained recommendation models, rather than requiring models to be trained from scratch. This makes fine-tuning techniques applicable even to deployed models that are typically trained extensively. 

\noindent \textbf{Should All Rank Lists be Stabilized with \method?}
\method fine-tunes a recommender to generate stable rank lists for all training instances. However, in some cases, the rank list is expected to change against perturbations. For instance, let us assume a cold-start user with very few interactions. If we perturb this user's interaction, the recommendation should be altered, since every single interaction of the cold-start user is crucial for its recommendation. Finding more types of rank lists not to stabilize is worth studying.

\noindent \textbf{Handling Diverse Perturbation Methods.}
In this paper, we focused on enhancing model stability against interaction-level perturbations such as injection, deletion, item replacement, and a mix of them. However, in the real world, there can be various types of perturbations such as user-, item-, or embedding-level perturbations. While \method can be easily extended to user- and item-level perturbations by performing the perturbation simulation at the user or item level, extending \method to embedding-level perturbations is worth investigating as finding embedding perturbations for our simulations is non-trivial.

\noindent \textbf{Extension to Non-Sequential Recommender Systems.}
As \method is optimized for sequential recommenders, its fine-tuning process should be modified for non-sequential recommendation models, such as collaborative filtering (CF). 
For instance, we can apply our rank-preserving regularization to each user instead of each training instance for CF-based recommenders.
\method can be generalized to multimodal recommendation setup, where recommenders employ additional modalities such as text or image features for training and predictions.
We leave the empirical validation of \method on such non-sequential recommenders as future work.

In conclusion, our work paves the path toward robust and reliable recommendation systems by proposing a novel fine-tuning method with perturbation simulations and rank-preserving regularization.
Future work includes extending \method to diverse recommendation models (e.g., reinforcement learning-based) and non-recommendation settings (e.g., information retrieval), other perturbation settings (e.g., embedding-level), and creating fine-tuning mechanisms for various content-aware recommendation models.
 
	\section*{Acknowledgments}
	{
	This research is supported in part by Georgia Institute of Technology, IDEaS, and Microsoft Azure. Sejoon Oh was partly supported by ML@GT, Twitch, and Kwanjeong fellowships.
	}	

	\bibliographystyle{ACM-Reference-Format}
	\bibliography{BIB/myref}

\end{document}